\documentclass{nature}
\usepackage{comment}
\usepackage{cite}
\usepackage[T1]{fontenc}
\usepackage[utf8]{inputenc}
\usepackage{graphicx}
\usepackage{float}
\usepackage{caption}
\usepackage{amssymb}
\usepackage{amsmath}
\usepackage{mathtools}
\usepackage{dcolumn}
\usepackage[final]{pdfpages}
\usepackage{color}
\usepackage{bm}
\usepackage[normalem]{ulem}
\usepackage{multibib}
\usepackage[section]{placeins}
\usepackage[labelfont=bf]{caption}
\usepackage{enumitem}  
\usepackage{chemformula}

\newcommand{\beq}{\begin{equation}}
\newcommand{\eeq}{\end{equation}}
\newcommand{\beqn}{\begin{eqnarray}}
\newcommand{\eeqn}{\end{eqnarray}}

\begin{document}

\title{Aharonov–Bohm Interference in Even-Denominator Fractional Quantum Hall States}
\author{Jehyun Kim$^{1,}$\footnote{These authors contributed equally to this work.}, Himanshu Dev$^{1,*}$, Amit Shaer$^1$, Ravi Kumar$^1$, Alexey Ilin$^1$, André Haug$^1$, Shelly Iskoz$^{1}$, Kenji Watanabe$^{2}$, Takashi Taniguchi$^{3}$, David F. Mross$^1$, Ady Stern$^1$ and Yuval Ronen$^{1}$\footnote{Corresponding author: yuval.ronen@weizmann.ac.il}}

\maketitle

\begin{affiliations}

\item Department of Condensed Matter Physics, Weizmann Institute of Science, Rehovot 76100, Israel.

\item Research Center for Functional Materials, National Institute for Materials Science, 1-1 Namiki, Tsukuba 305-0044, Japan.
\item International Center for Materials Nanoarchitectonics, National Institute for Materials Science, 1-1 Namiki, Tsukuba 305-0044, Japan.
\\

\end{affiliations}

\begin{abstract}

Position exchange of non-Abelian anyons affects the quantum state of their system in a topologically-protected way. Their expected manifestations in even-denominator fractional quantum Hall (FQH) systems offer the opportunity to directly study their unique statistical properties in interference experiments. In this work, we present the observation of coherent Aharonov–Bohm interference at two even-denominator states in high-mobility bilayer graphene-based van der Waals heterostructures by employing the Fabry–Pérot interferometry (FPI) technique. Operating the interferometer at a constant filling factor, we observe an oscillation period corresponding to two flux quanta inside the interference loop, $\Delta\Phi=2\Phi_0$, at which the interference does not carry signatures of non-Abelian statistics. The absence of the expected periodicity of $\Delta\Phi=4\Phi_0$ may indicate that the interfering quasiparticles carry the charge $e^* = \frac{1}{2}e$ or that interference of $e^* = \frac{1}{4}e$ quasiparticles is thermally smeared. Interestingly, at two hole-conjugate states, we also observe oscillation periods of half the expected value, indicating interference of $e^* = \frac{2}{3}e$ quasiparticles instead of $e^* = \frac{1}{3}e$. To probe statistical phase contributions, we operated the FPI with controlled deviations of the filling factor, thereby introducing fractional quasiparticles inside the interference loop. The resulting changes to the interference patterns at both half-filled states indicate that the additional bulk quasiparticles carry the fundamental charge $e^*=\frac{1}{4}e$, as expected for non-Abelian anyons.

\end{abstract}

\clearpage
\noindent\textbf{Introduction}

For over four decades, quasiparticles carrying fractional charge and obeying fractional statistics have captivated the condensed matter physics community.\cite{feldman2021fractional} Their most prevalent types are the Abelian anyons, which exhibit quantized exchange phases lying in between those of bosons and fermions. Even more remarkable are non-Abelian anyons that can fundamentally transform the many-body wavefunction through particle exchanges, processing quantum information in a topologically protected manner.\cite{nayak2008non} The fractional quantum Hall (FQH) systems have emerged as a leading platform for realizing and manipulating these exotic quasiparticles, owing to high electron mobility, long coherence times, and exceptional controllability.\cite{du2009fractional} Fractional charge was first observed via shot noise measurements at odd-denominator filling factors expected to host Abelian states,\cite{saminadayar1997observation, de1998direct} and later at the even-denominator filling $\nu=\frac{5}{2}$ in GaAs,\cite{dolev2008observation, venkatachalam2011local} a leading candidate for non-Abelian topological order.\cite{MA2024324}

Direct measurements of anyonic exchange statistics require phase-sensitive techniques such as quantum Hall interferometry in the Aharonov-Bohm (AB) regime, where Coulomb interactions are sufficiently weak for the interferometer area to remain constant as $B$ is varied.\cite{halperin2011theory} Seminal works by Nakamura \textit{et al.} demonstrated AB interference of fractionally-charged quasiparticles using a GaAs Fabry-P\'erot Interferometer (FPI) at filling $\nu=\frac{1}{3}$,\cite{nakamura2019aharonov} and braiding (double-exchange) phases in a subsequent study.\cite{nakamura2020direct} These findings were generalized to different filling factors,\cite{nakamura2023fabry} platforms,\cite{kim2024aharonov, werkmeister2024anyon, samuelson2024anyonic} and interferometer architectures.\cite{ghosh2024anyonic} In parallel, time-domain braiding experiments \cite{bartolomei2020fractional,taktak2022two, ruelle2023comparing, lee2023partitioning,glidic2023cross,glidic2024signature} also support anyonic quasiparticle statistics in Abelian FQH states.\cite{rosenow2016current} At even-denominator fillings, FPI studies at $\nu=\frac{5}{2}$ in GaAs have reported signatures consistent with non-Abelian statistics.\cite{willett2023interference} However, the interpretation of those experiments remains challenging, primarily due to the absence of robust AB interference.

Even-denominator states have been observed in several FQH platforms, including GaAs,\cite{willett1987observation} ZnO,\cite{falson2015even} graphene,\cite{ narayanan2018incompressible, kim2019even } bilayer graphene,\cite{ki2014observation,li2017even, zibrov2017tunable, zibrov2018even, kumar2024quarter} and WSe$_{2}$.\cite{shi2020odd} In GaAs narrow quantum wells, thermal transport mea\-sure\-ments\cite{banerjee2018observation, dutta2022distinguishing} consistently support a non-Abelian topological order known as PH-Pfaffian.\cite{son2015composite} Distinct non-Abelian orders known as Moore-Read Pfaffian\cite{moore1991nonabelions} and anti-Pfaffian\cite{lee2007particle, levin2007particle} are indicated by daughter states\cite{levin2009collective, yutushui2024paired, zheltonozhskii2024identifying, zhang2024hierarchy} in bilayer graphene and GaAs wide quantum wells. Specifically, bilayer graphene realizes quantized plateaus at seven half-integer filling factors in the zeroth Landau level. Moreover, the presumed topological orders alternate between Pfaffian and anti-Pfaffian, offering a rich playground for interference studies of non-Abelian anyons.

In this work, we report the observation of robust Aharonov–Bohm oscillations at two even-denominator FQH plateaus in bilayer graphene. Employing a gate-defined FPI in a high-mobility bilayer graphene-based van der Waals (vdW) heterostructures, we perform a detailed study of the interference patterns as a function of magnetic field, area, and density. At both fillings, we observe the unexpected AB periodicity $\Delta \Phi = 2\Phi_0$ when the magnetic field and the density are varied together to maintain contact filling. The most conservative interpretation of these measurements is the interference of quasiparticles with charge $e^*=\frac{1}{2}e$, twice the charge expected theoretically \cite{moore1991nonabelions} and observed in earlier shot noise and SET measurements in GaAs.\cite{dolev2008observation, venkatachalam2011local} However, this frequency could also originate from $e^*=\frac{1}{4}e$ quasiparticles performing an even number of loops. 

This finding prompted us to study the nearby odd-denominator states at Landau-level fillings of $\nu=\frac{1}{3},\frac{2}{3}$, where we found AB periodicities corresponding to interference of quasiparticles with charges $e^*=\frac{1}{3}e$ and $e^*=\frac{2}{3}e$, respectively. Across the three fillings, the interfering charge follows $e^* = \nu e$ instead of the minimal charges of bulk quasiparticles, which are $\frac{1}{3}e,\frac{1}{4}e,\frac{1}{3}e$ for these states. We note that in GaAs, shot-noise measurements at hole conjugate states also find a partitioned charge of $\nu e$,\cite{bid2009shot, bhattacharyya2019melting} but interference at $\nu=\frac{2}{3}$ shows $e^*=e$.\cite{nakamura2019aharonov} Finally, by tuning the electron density independently of the magnetic field, we deviate from the fixed-filling constraint, thereby introducing localized bulk quasiparticles.\cite{arovas1984fractional} Unlike the integer case, we observe a statistical contribution to the interference phase of fractional fillings, supporting their anyonic character.

\noindent\textbf{Design and measurement phase-space of a bilayer-graphene-based FPI}

The FPI device is constructed on a high-mobility vdW heterostructure, with bilayer graphene as the active two-dimensional layer, which is encapsulated between hexagonal boron nitride dielectric layers, while conductive graphite layers on the top and bottom serve as gates. The heterostructure design and nanofabrication techniques follow those detailed in our previous study,\cite{kim2024aharonov} with measurements conducted under a perpendicular magnetic field up to $B=12$ T and at a base temperature of $T=10$ mK.

\begin{figure}[h]
  \centering
  \includegraphics[width=.9\textwidth]{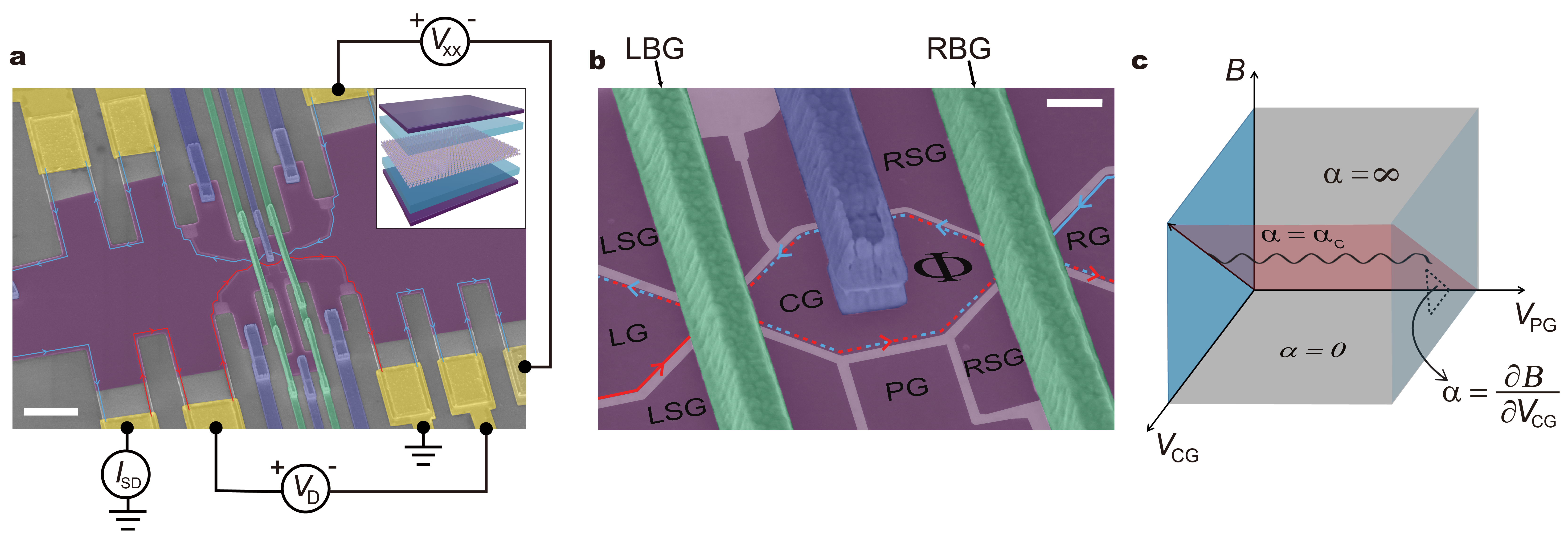}
  \caption{\textbf{Fabry-Pérot Interferometer (FPI) based on the bilayer graphene.} (\textbf{a}) Measurement configurations depicted as a false-color scanning electron microscopy image of the FPI in a bilayer graphene heterostructure (see inset). The top graphite layer (purple) is divided into eight separate regions via etched trenches, better seen in (b). Each region acts as a gate, electrostatically tuned via air bridges (blue) to define the interferometer. The device dimensions are indicated by the white scale bar of length 3 $\mu$m. In the quantum Hall regime, current $I_\mathrm{SD}$ applied through an ohmic contact (yellow) propagates via edge modes and is partitioned by two quantum point contacts (QPCs) formed by the left and right split gates (LSG and RSG), resulting in oscillating diagonal resistance $R_\mathrm{D}=(V_\mathrm{D}\textsuperscript{+}-V_\mathrm{D}\textsuperscript{-})/I_\mathrm{SD}$. (\textbf{b}) Magnification of the interfering region near the center gate (CG). Left and right air bridges (LBG and RBG, shown in green) are suspended 200 nm above each QPC region, fine-tuning the saddle point potential. The lithographic interference area, determined by CG area, is 1 $\mu$$\mathrm{m}^2$. (\textbf{c}) Measurement phase space defined by $B$, $V_\mathrm{PG}$, and $V_\mathrm{CG}$. $R_\mathrm{D}$ is measured along planes defined by $\alpha=\frac{\partial B}{\partial V_\mathrm{CG}}$ and displayed as two-dimensional $B|_\alpha$-$V_\mathrm{PG}$ pajamas.} 
  \label{fig1}
\end{figure}

A false-color scanning electron microscopy image of the FPI is shown in Fig.~\ref{fig1}a. The top graphite layer is divided into eight distinct regions by 40nm wide etched trenches, with each region contacted via air bridges. Together with a global graphite back gate (BG), these eight top gates enable capacitive tuning of the potential and displacement fields across various regions of the bilayer graphene. The filling factor inside the interferometer is controlled by the center gate (CG), while the left and right gates (LG and RG) set the outer fillings. The two quantum point contacts (QPCs) are formed by the split gates LSG and RSG, which set the filling underneath to zero, thereby guiding the counter-propagating edge modes on opposite sides into close proximity and introducing tunneling between them. An additional plunger gate (PG) allows fine control over the area enclosed by the interfering QH edge mode. Fig.~\ref{fig1}b provides a zoomed-in view of the interfering region, lithographically defined to be 1 $\mu\mathrm{m}^2$(see SI1). Two air bridges, LBG and RBG, positioned 200 nm above the QPC regions, act as gates, fine-tuning the transmission of each QPC independently. 

We inject a bias current $I_\mathrm{SD}$, which propagates along the FQH edge modes with an anti-clockwise (clockwise) chirality for electron (hole) carriers, impinging on the FPI as illustrated in Fig.~\ref{fig1}a. Current is collected on the other side of the interferometer by a single ground, while measuring the diagonal resistance $R_\mathrm{D}=\frac{V_\mathrm{D}\textsuperscript{+}-V_\mathrm{D}\textsuperscript{-}}{I_\mathrm{SD}}$ to reveal interference. In the low backscattering regime, $R_\mathrm{D}$ includes an oscillatory contribution $\Delta R_\mathrm{D} \propto \cos~\theta$, where the interference phase $\theta$ is composed of both AB and statistical phases, i.e.,\cite{ kivelson1990semiclassical, chamon1997two, halperin2011theory }
\begin{equation} \label{eq:1}  
\theta = \theta_\text{AB} + \theta_\text{stat} = 2\pi \frac{e^*}{e} \frac{ A B }{\Phi_0} + N_\mathrm{qp}\theta_\mathrm{anyon} ~,
\tag{1}
\end{equation}
where $A$ is the interfering area, $N_\mathrm{qp}$ is the integer number of localized QPs within the interference loop, and $\theta_\mathrm{anyon}$ the braiding phase. For non-Abelian quasiparticles, $R_\mathrm{D}$ is predicted to follow a more intricate pattern that differs for even and odd $N_\mathrm{qp}$.\cite{stern2006proposed, bonderson2006detecting }

We perform measurements of $R_\mathrm{D}$ in the three-dimensional parameter space spanned by the magnetic field $B$, the PG voltage $V_\mathrm{PG}$, and the CG voltage $V_\mathrm{CG}$; see Fig.~\ref{fig1}c. To disentangle the two terms in $\theta$ of Eq.~\eqref{eq:1}, we follow lines of different slopes $\alpha=\frac{\partial B}{\partial V_\mathrm{CG}}$ in the $B$--$V_\text{CG}$ plane. 
The AB contribution is isolated at the critical trajectories $\alpha_c$, for which charges are continuously added to the interference loop to maintain constant fillings.

Consequently, $R_\mathrm{D}$ follows the well-known "pajama pattern" in the $B$-$V_\mathrm{PG}$ plane with a flux periodicity set by the interfering quasiparticle charge $e^*$. Deviations from this trajectory introduce bulk quasiparticles, $N_\text{qp}$, which are expected to manifest individually via phase slips, and which alter the average flux periodicity. Other significant trajectories include constant density, $\alpha=\infty$, and constant magnetic field, $\alpha=0$, illustrated in Fig.~\ref{fig1}c.

\noindent\textbf{Even-denominator Aharonov–Bohm interference}

\begin{figure}[H]
  \centering
  \includegraphics[width=.95\textwidth]{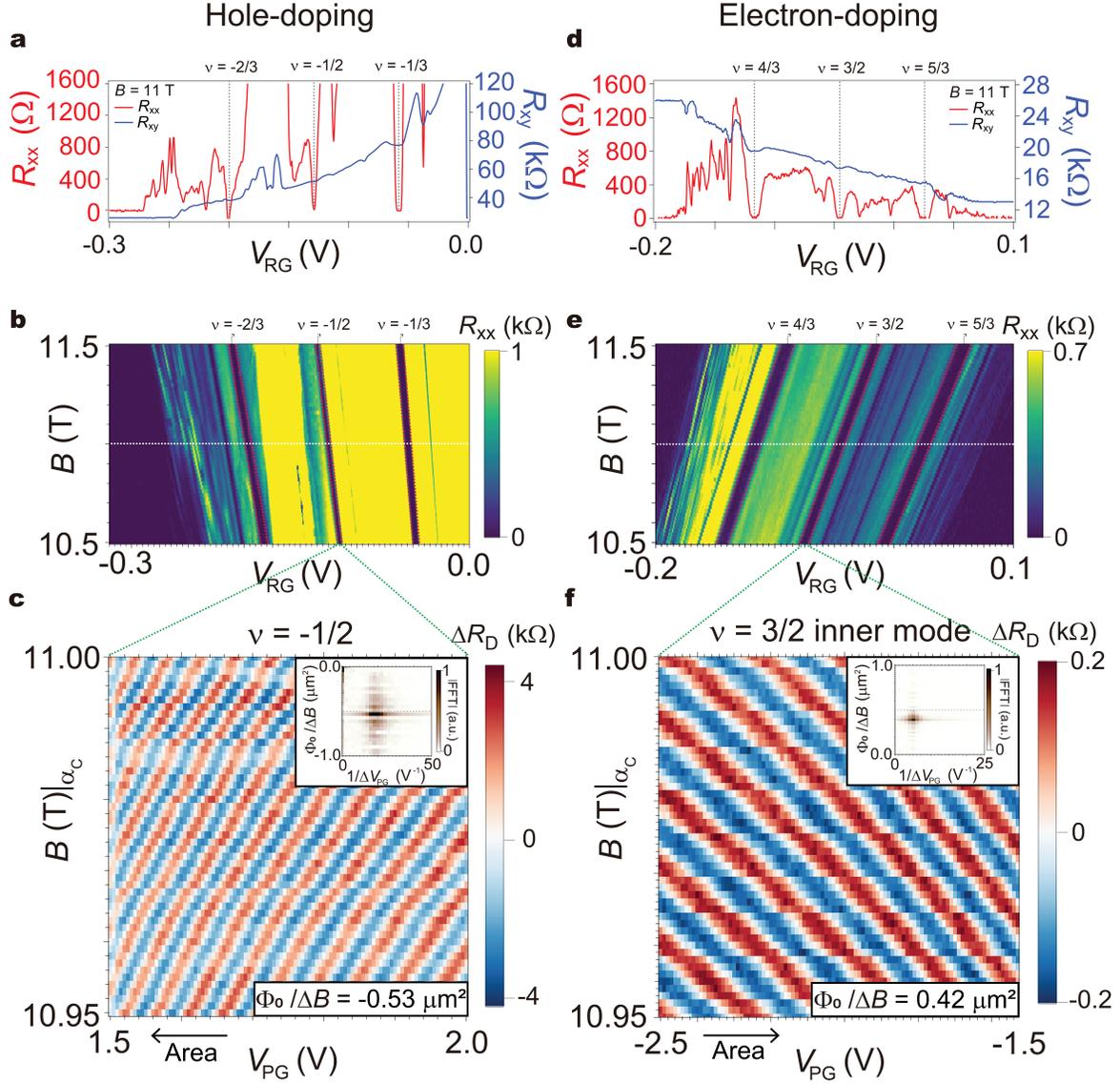}
  \caption{\textbf{Even-denominator Aharonov-Bohm interference.} (\textbf{a}) Longitudinal resistance $R_\mathrm{xx}$ and Hall resistance $R_\mathrm{xy}$ measured at 11 T on the right side of the FPI, clearly showing fully developed even and odd denominator quantum Hall states at $\nu=-\frac{2}{3},-\frac{1}{2}$, and $-\frac{1}{3}$ (\textbf{b}) $R_\mathrm{xx}$ fan diagram performed on the right side of the FPI between 10.5 and 11.5 T. Red dashed lines indicate the boundaries for each quantum Hall state (\textbf{c}) $\Delta$$R_\mathrm{D}$ at $\nu=-\frac{1}{2}$ displayed as a $B|_{\alpha_{c}}$-$V_\text{PG}$ pajama plot, showing clear AB oscillations. Inset: 2D-FFT analysis used to extract the magnetic field periodicity $\frac{\Phi_0}{\Delta B}$ shown on the right lower side of the pajama. (\textbf{d},\textbf{e}) same as (a,b) for the electron-doped filling factors $\nu=\frac{4}{3},\frac{3}{2}$, and $\frac{5}{3}$. (\textbf{f}) Same as (c) for $\nu=\frac{3}{2}$ with partitioning of the fractional inner mode.}
  \label{fig2}
\end{figure}

We begin the FPI study of even denominator states at the filling factor $\nu=-\frac{1}{2}$ due to its simple edge structure, which consists only of fractional modes. Fig.~\ref{fig2}a displays the longitudinal resistance $R_\mathrm{xx}$ and Hall resistance $R_\mathrm{xy}$ measured at 11 T on the right side of the FPI. The data clearly reveal fully developed integer and fractional QH states at $\nu = -1$, $-\frac{2}{3}$, $-\frac{1}{2}$, and $-\frac{1}{3}$. Fig.~\ref{fig2}b presents an $R_\mathrm{xx}$ fan diagram, which we use to extract the constant-filling factor trajectories. We define $\alpha_c = \frac{\Phi_0}{\nu e} C$, with $C=\frac{1}{A}\frac{d Q}{ d V_\text{RG}}$ the capacitance per unit area between the right gate and the bilayer graphene underneath, extracted from the Streda formula for each fractional state as the center of the incompressible region, whose boundaries are indicated by red dashed lines (see SI2).

Focusing on the $-\frac{1}{2}$ state, Fig.~\ref{fig2}c shows the interference pattern as a function of $V_\text{PG}$ and $B|_{\alpha_{c}}$, where the $\alpha_c$ constraint indicates that $V_\text{CG}$ is adjusted to maintain constant filling. Specifically, we present the data as $\Delta R_\mathrm{D}$=$R_\mathrm{D}-\langle R_\mathrm{D}\rangle$, subtracting the average value at each magnetic field. The positive slope of the pajama indicates AB-dominated interference, since increasing $V_\text{PG}$ decreases the interference area for hole-doped states. The measured visibility is around 1.9\%, comparable to the one at integers and odd-denominator states (See SI4). To extract the flux periodicity, we perform a 2D fast Fourier transform (2D-FFT), shown in the inset of Fig.~\ref{fig2}c as a function of $\frac{\Phi_0}{\Delta B}$ and $\frac{1}{\Delta V_\text{PG}}$. From the magnetic field periodicity, we extract $ A\frac{\Phi_0}{ \Delta \Phi} \approx -0.53$ $\mu$m$^2$. The lithographic area $A \approx 1$ $\mu$m$^2$ agrees with the one extracted from interference at $\nu=-1$ to within 2\% (see SI6). Using the same area at $\nu=-\frac{1}{2}$ yields the unexpected flux periodicity $\Delta \Phi = (1.89 \pm 0.26) \Phi_0 \approx 2 \Phi_0$.

 Following the first term in Eq.~\eqref{eq:1}, this periodicity suggests an interfering quasiparticle charge of $e^* = \frac{1}{2}e$, which tunnels across the QPCs to form an interference loop. Quasiparticles with this charge exist as bulk excitations at half-filling, arising from the fusion of two fundamental quasiparticles carrying charge $\frac{1}{4}e$, in all Abelian or non-Abelian FQH candidate states. Alternatively, this periodicity could also arise in a scenario where non-Abelian $\frac{1}{4}e$ quasiparticles interfere. In that case, when a non-zero number of non-Abelian quasiparticles are localized in the bulk, there are multiple degenerate ground states. Fluctuations between these ground states on the time scale of the measurement could suppress the $4\Phi_0$ periodicity that arises from a single winding of $\frac{1}{4}e$ quasiparticles while not affecting the $2\Phi_0$ periodicity arising from double windings or $\frac{1}{2}e$ quasiparticles.\cite{stern2006proposed,bonderson2006detecting}

\clearpage
To test the generality of these findings, we investigated the $\nu=\frac{3}{2}$ plateau (on the electron side), which exhibits a gap comparable to $\nu=-\frac{1}{2}$.\cite{li2017even,zibrov2017tunable} Similar to the previous case, Fig.~\ref{fig2}d displays $R_\mathrm{xx}$ and $R_\mathrm{xy}$ measured at 11 T, revealing well-developed FQH states at $\nu = \frac{4}{3}$, $\frac{3}{2}$, and $\frac{5}{3}$. Fig.~\ref{fig2}e presents an $R_\mathrm{xx}$ fan diagram, which we use to extract $\alpha_c$ as before. Fig.~\ref{fig2}f shows the interference pattern as a function of $V_\text{PG}$ and $B|_{\alpha_c}$ that arises when the QPCs are tuned to partition the fractional inner edge mode (see SI10) for the interference of the integer outer edge). The slope of the pajama pattern with 5.6\% visibility is opposite to $\nu=-\frac{1}{2}$, indicating AB-dominated interference for electron-doped states (see SI3). The 2D-FFT, shown in the inset of Fig.~\ref{fig2}f, yields the magnetic field periodicity $A\frac{\Phi_0}{ \Delta \Phi} \approx 0.42$ $\mu$m$^2$. Estimating the interfering area based on the integers $\nu=1,2$, we find $A=0.99\pm0.10 \mu \mathrm{m}^2$, consistent with the lithographic area (see SI7). Using this area at $\nu=\frac{3}{2}$, we conclude $\Delta \Phi =(2.35 \pm 0.78) \Phi_0\approx 2 \Phi_0$. Temperature dependence measurements showed a reduction in visibility with increasing temperature, while the magnetic field periodicity remained constant (see SI8 and SI9).

These two measurements consistently show periodicities close to $2\Phi_0$ and not the expected $4\Phi_0$. The observations reflect the interference of $\frac{1}{2}e$ quasiparticles at $\nu=-\frac{1}{2}$ and $\frac{3}{2}$. We note that the topological orders at both fillings are believed to be Pfaffian,\cite{huang2022valley,hu2023high,assouline2024energy,kumar2024quarter} but their edge structures at the boundary to $\nu=0$ are qualitatively different. In particular, a Pfaffian order at $\nu=-\frac{1}{2}$ would exhibit an anti-Pfaffian edge state with three upstream Majorana fermions. Insofar as the identification of these states is accurate, our experiment effectively probes two distinct non-Abelian topological orders.

\noindent\textbf{Interference of} $e^* = \nu_\text{LL} e$ \textbf{quasiparticles in various FQH states}

\begin{figure}[H]
  \centering
  \includegraphics[width=.95\textwidth]{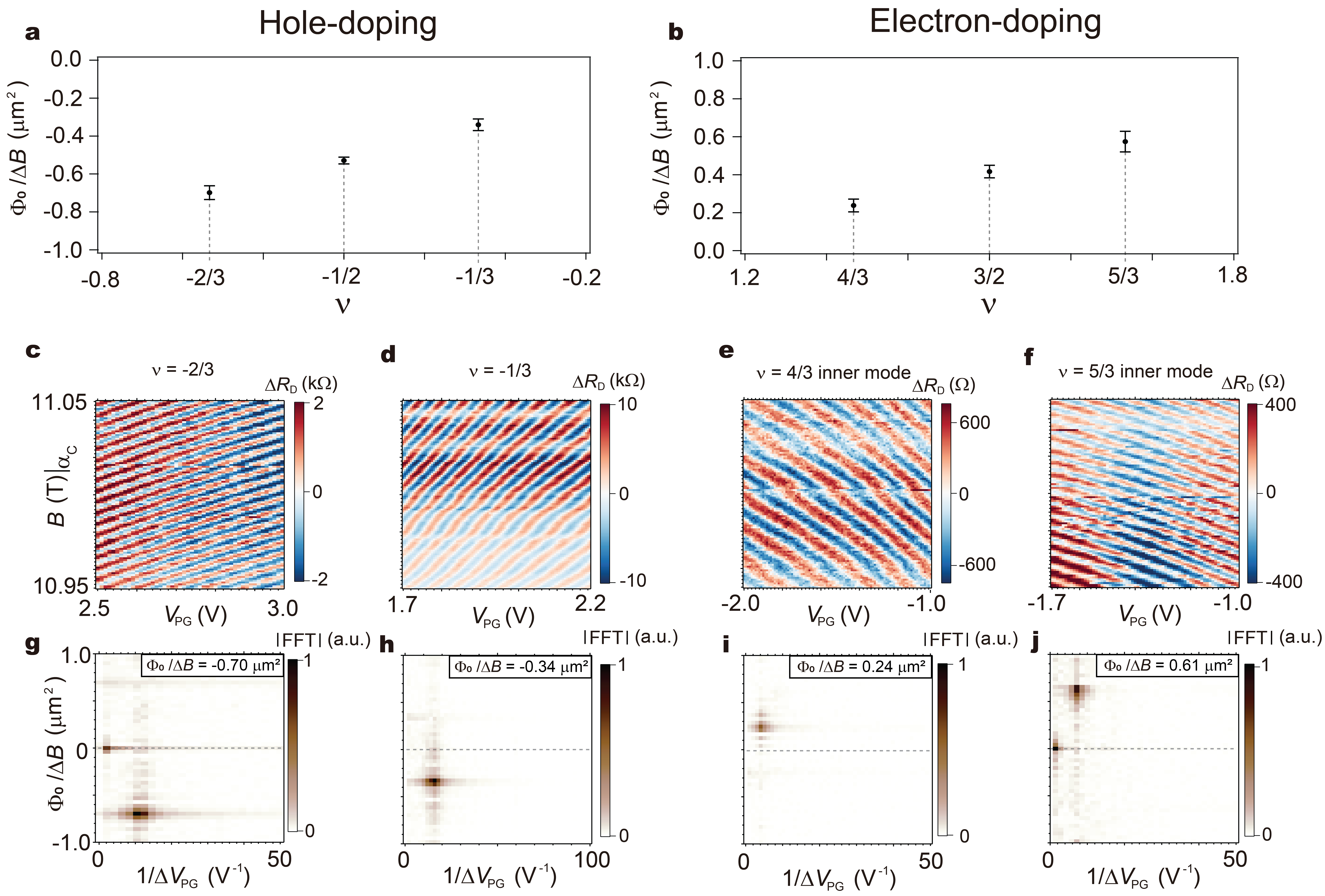}
  \caption{\textbf{Interference of} $e^* = \nu_\text{LL} e$ \textbf{quasiparticles in various FQH states.} (\textbf{a}) Magnetic field periodicities $\frac{\Phi_0}{\Delta B}$ at constant filling extracted from 2D-FFT analyses at $\nu=-\frac{2}{3}$, $-\frac{1}{2}$, and $-\frac{1}{3}$. (\textbf{b}) Same for $\nu=\frac{4}{3}$, $\frac{3}{2}$, and $\frac{5}{3}$ with partitioning of the fractional inner mode. 
  (\textbf{c-f}) $\Delta$$R_\mathrm{D}$ shown as $V_\text{PG}$-$B|_{\alpha_{c}}$ pajamas for $\nu=-\frac{2}{3}$, $-\frac{1}{3}$, $\nu=\frac{4}{3}$, and $\frac{5}{3}$, respectively. At $\nu=\frac{5}{3}$, $\alpha$ deviated from $\alpha_c$ by $3\%$. (\textbf{g-j}) Corresponding 2D-FFTs, with the extracted magnetic field periodicities in the insets. }
  \label{fig3}
\end{figure}

At both even-denominator filling factors, the observed Aharonov-Bohm periodicity is consistent with an interfering quasiparticle charge that matches the Landau-level filling factor $\nu_\text{LL}=\frac{1}{2}$. The interfering charge at $\nu=\frac{1}{3}$ also follows the filling factor.\cite{nakamura2019aharonov,werkmeister2024anyon,samuelson2024anyonic} We extended the study to hole-conjugate states at $\nu=-\frac{2}{3}$ and $\frac{5}{3}$ to determine if their interfering charge is also set by the filling or by the minimal bulk excitation. The particle-like states at $\nu=-\frac{1}{3}$ and $\frac{4}{3}$ were also included as known reference points. Figs.~\ref{fig3}a and 3b show the extracted flux periodicities for all six fractional fillings in constant filling measurements. The values for the odd-denominators are extracted from the pajama patterns in Fig.~\ref{fig3}c-f via the 2D FFTs shown in Fig.~\ref{fig3}g-j, assuming the same interference areas for hole-doped and electron-doped states as before. The results confirm the interference of $e^* = \nu_\text{LL} e$ quasiparticles in all states included in our study. A recent experiment on the hole conjugate states $\nu=\frac{2}{3},\frac{3}{5}$ and $\frac{4}{7}$ in GaAs using a Mach-Zehnder interferometer also observed interference of $e^* = \nu_\text{LL} e$ quasiparticles.\cite{ghosh2024bunching}
It is not understood why non-fundamental quasiparticles should dominate the interference, as our measurements at half-filled and hole-conjugate states indicate. We point out that previous interference experiments at the hole-conjugate $\nu=\frac{2}{3}$ state in GaAs reported the periodicity $\Delta \Phi = \Phi_0$ corresponding to the interference of electrons.\cite{nakamura2019aharonov} Moreover, Mach-Zehnder interference of the higher particle-like Jain states $\nu=\frac{2}{5},\frac{3}{7}$ observed $\Delta \Phi = 5\Phi_0,7\Phi_0$,\cite{ghosh2024anyonic} corresponding to the fundamental quasiparticle charge instead of $ \nu_\text{LL} e$.

Theoretically, the question of which type of quasiparticle tunnels is addressed based on the renormalization of bare tunneling amplitudes by the interactions intrinsic to fractional edge modes. The bare tunneling amplitudes for different quasiparticles are non-universal and hard to calculate reliably. Their renormalization, encoded via a scaling dimension of tunneling operators, is the same for fundamental and $e^* =\frac{2}{3} e$ quasiparticles at the $\nu=\frac{2}{3}$ edge.\cite{kane1994randomness} It is possible that interactions across the QPC tip the balance in favor of $e^* =\frac{2}{3} e$ tunneling. Alternatively, when both tunneling processes occur with comparable probabilities, the $3 \Phi_0$ periodicity expected for $\frac{1}{3}e$ quasiparticles could be thermally suppressed because it requires exciting a neutral mode, which propagates with a much smaller velocity than the charge mode. At half-filling, the scaling dimensions of $e^* =\frac{1}{4} e$ quasiparticles depend on which topological state is realized, but their numerical values are generally close to those of $e^* =\frac{1}{2}e$ tunneling. Different interactions across the QPC could favor either one, and a similar thermal suppression may affect the $\frac{1}{4}e$ quasiparticle, which also excites a neutral mode. We also caution that the scaling dimensions extracted from experiments often deviate significantly from theoretical expectations.\cite{chang2003chiral, radu2008quasi, lin2012measurements}

\noindent\textbf{The statistical contribution to the interference of fractional quasiparticles}

Interference of fractional quasiparticles fundamentally differs from that of electrons by quantum statistical effects, i.e., the second term in Eq.~\eqref{eq:1}. Interfering quasiparticles acquire a quantized phase change for each localized anyon in the interferometer bulk. For electron interference, this additional phase is an unobservable multiple of $2\pi$ independent of the bulk anyon type. To observe such contributions, we operate the FPI at $\alpha \neq \alpha_c$, such that tuning the magnetic field or $V_\text{CG}$ causes the filling factor to deviate from its rational value ($\nu=\frac{p}{q}$), which introduces excess charge carriers in the form of quasiparticles inside the interference loop. Each well-isolated quasiparticle in the bulk is expected to result in a sharp phase jump in the interference pattern.\cite{nakamura2020direct,kim2024aharonov,samuelson2024anyonic} Consequently, introducing quasiparticles at a constant rate along a fixed $\alpha$ trajectory alters the overall slope of the constant-phase lines in the pajama pattern.

The change of the slope provides crucial insights into which quasiparticles enter the interference loop as the filling factor varies. For the case where interfering quasiparticles entering Eq.~\eqref{eq:1} carry charge $e^* = \nu_\text{LL} e$ and fundamental quasiparticles are introduced into the bulk, we find 
\begin{equation} \label{eq:2} 
\begin{split}
& \text{Integer edge modes}:\qquad\qquad \ \ \ \ \ \frac{\Phi_0}{\Delta B}=A = \text{const} ~,\\
& \text{Fractional edge modes}:\qquad\qquad  \frac{\Phi_0}{\Delta B}= \left(\nu_{LL}-\nu\right)A + \nu \frac{\alpha_c}{\alpha}A ~.
 \end{split}\tag{2}
\end{equation}
 Since the mutual statistics with the $e^* = \frac{1}{2}e$ quasiparticles with all other quasiparticles is Abelian, Eq.~\eqref{eq:2} holds for all paired states. The second line matches the phase $\theta=2\pi \langle N\rangle$ with $N$ the number of electrons in the loop, generalizing the result by Arovas et al.\cite{arovas1984fractional} If filling factor deviations introduce quasiparticles other than the fundamental ones into the bulk, the slope on the right-hand side changes (see SI11).

\begin{figure}[h]
  \centering
\centerline{\includegraphics[width=.85\textwidth]{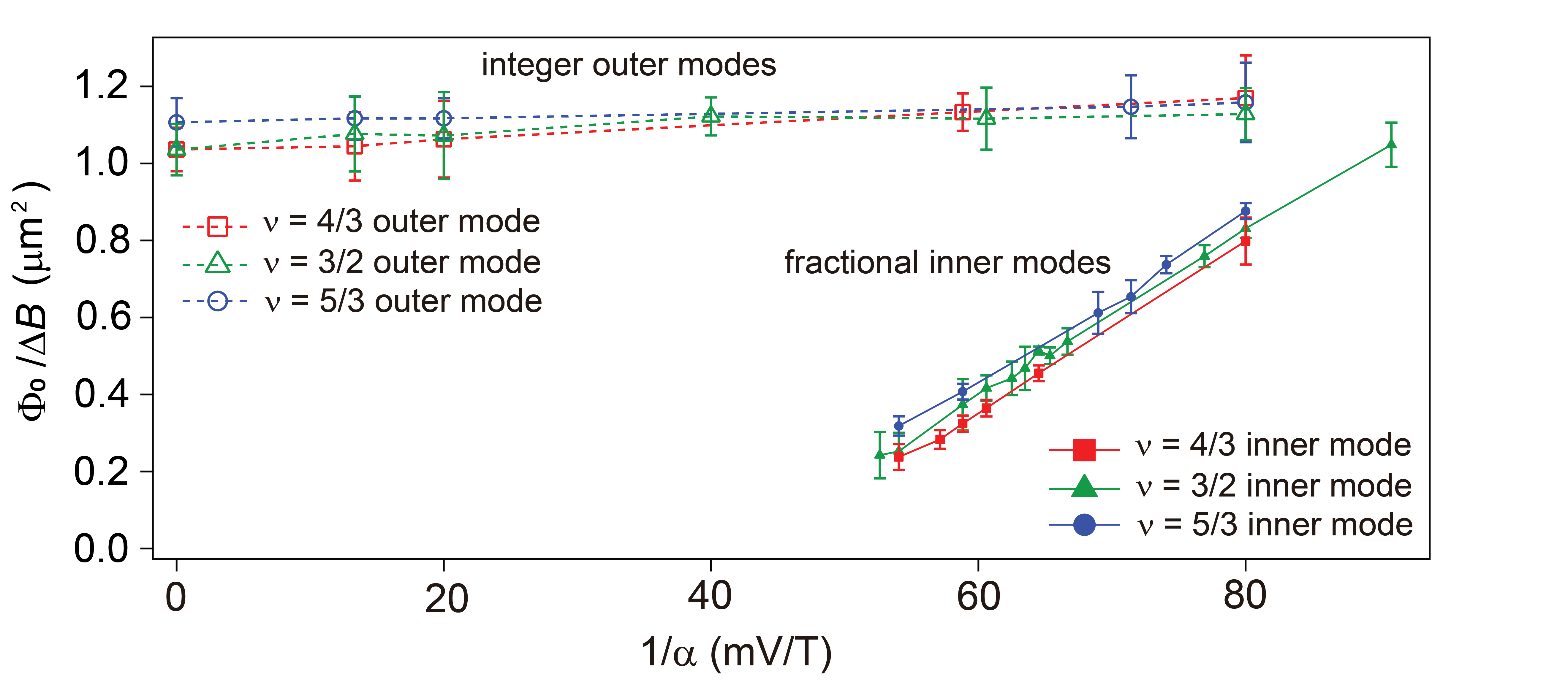}}
  \caption{\textbf{The statistical contribution to the interference of fractional quasiparticles.} Magnetic field periodicities $\frac{\Phi_0}{\Delta B}$ obtained along different trajectories $\alpha$ from 2D-FFTs for the fractional inner and integer outer modes of $\nu=\frac{4}{3}$, $\frac{3}{2}$, and $\frac{5}{3}$.}
  \label{fig4}
\end{figure}

We extract $\frac{\Phi_0}{\Delta B}$ from the 2D FFT for the fillings $\nu=\frac{4}{3},\frac{3}{2}$ and $\frac{5}{3}$, both for partitioning of the fractional inner modes and integer outer modes. Fig.~\ref{fig4} shows our results for each $\alpha$, with $\frac{\Phi_0}{\Delta B}$ obtained from the 2D-FFT of the corresponding pajama patterns. For all the integer outer modes, $\Delta B$ is independent of the $\alpha$, as expected. In contrast, for the fractional modes, all measurements collapse into a single linear dependence on $\frac{1}{\alpha}$ as in the second line of Eq.~\eqref{eq:2}. Their slope deviates by some 15\% from the numerical value expected based on the bulk capacitance $C$, obtained via the Streda formula for the region to the right of the FPI. This discrepancy can arise from boundary effects of the comparatively small center gate, small changes in the interference area with $V_\text{CG}$, and bulk-edge couplings (see SI12).

At $\nu=\frac{3}{2}$ with an interfering inner mode, the observed slope confirms the predicted statistical contribution to the interference of $e^*=\frac{1}{2}e$ quasiparticles, which are Abelian. In particular, this measurement indicates that quarter-charge bulk quasiparticles are introduced as the filling factor deviates from half-filling and not half-charged ones.

\noindent\textbf{Interference patterns at a constant magnetic field ($\alpha$ = 0)}

\begin{figure}[H]
  \centering
  \centerline{\includegraphics[width=.95\textwidth]{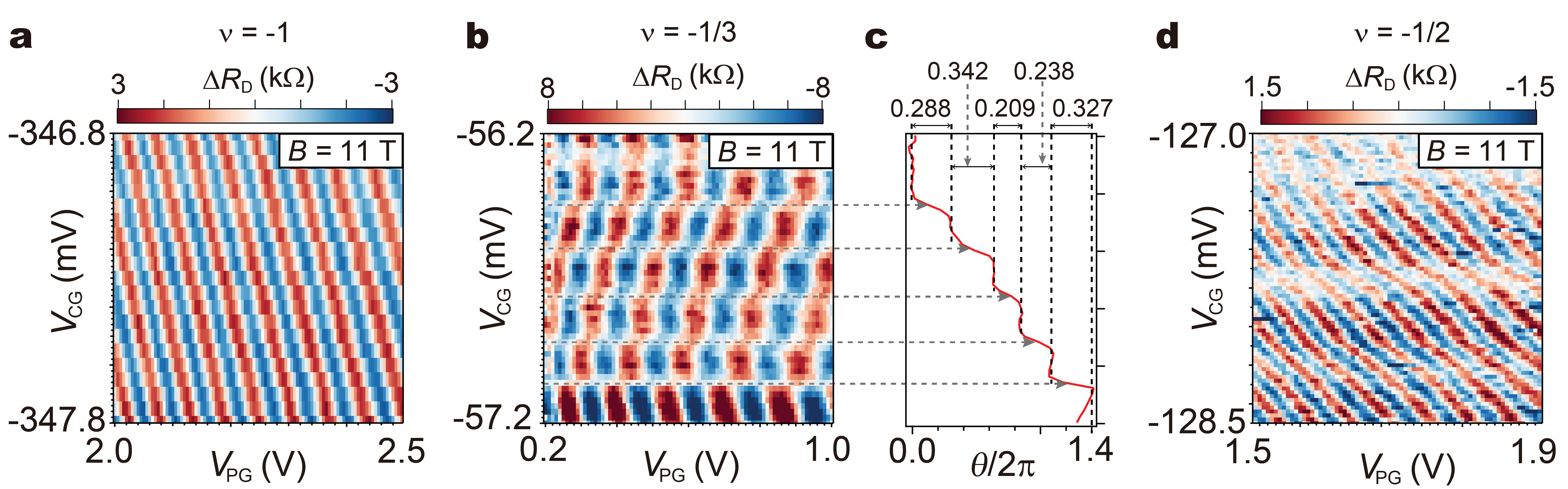}}
  \caption{\textbf{Interference patterns at a constant magnetic field ($\alpha$ = 0).} (\textbf{a}) $\Delta R_\mathrm{D}$ shown as a $V_\text{CG}$-$V_\text{PG}$ pajama, measured in a constant magnetic field plane at $\nu = -1$. (\textbf{b}) The $V_\text{PG}$-$V_\text{CG}$ pajama at $\nu=-\frac{1}{3}$ contains clear phase jumps approximately equally spaced in $V_\text{CG}$ (\textbf{c}) Phase slips extracted from 1D-FFTs at each $V_\text{CG}$, subtracting the local AB contribution. (\textbf{d}) $V_\text{CG}$-$V_\text{PG}$ pajama at $\nu=-\frac{1}{2}$.}
  \label{fig5}
\end{figure}

Lastly, we measured $R_\mathrm{D}$ as a function of $V_\mathrm{CG}$ and $V_\mathrm{PG}$ at a constant $B=11$ T, \textit{i.e.}, the $\alpha=0$ trajectory. Ideally, in this regime, the interference pattern would be influenced solely by the statistical term. However, variations in the area induced by changes in $V_\mathrm{CG}$ and $V_\mathrm{PG}$ introduce an AB contribution, as can be seen in the $\nu=-1$ pajama of Fig.~\ref{fig5}a. As expected for an integer state, it does not contain phase jumps. In contrast, the pajama pattern for $\nu=-\frac{1}{3}$, shown in Fig.~\ref{fig5}b, exhibits clear phase jumps. The magnitude of these jumps, extracted via the 1D-FFT analysis shown in Fig.~\ref{fig5}c, is close to the theoretically expected value $\theta_\text{anyon}=\frac{2\pi}{3}$ (see SI5). Fig.~\ref{fig5}d shows a representative pajama pattern for $\nu=-\frac{1}{2}$. Here, the phase variations in the pajama pattern are more pronounced, and we cannot identify clear phase jumps.

\noindent\textbf{Conclusions}

Our results mark two significant advancements towards the long-standing goal of observing non-Abelian statistics, bringing this objective within reach. The first essential condition---quasiparticle coherence in candidate non-Abelian states---is demonstrated by the observation of Aharonov–Bohm interference at two even-denominator FQH states. The second condition---interference contributions from localized bulk quasiparticles---was also observed by tuning the magnetic field and density to deviate from constant filling. Notably, our observations indicate that the localized bulk quasiparticles exhibit a charge of $\frac{1}{4}e$, as expected for non-Abelian quasiparticles.

The observed flux periodicity of $2 \Phi_0$ is consistent with the non-Abelian double-winding scenario or with a scenario where it arises from the interference of Abelian $\frac{1}{2}e$ quasiparticles. Distinguishing between these two possibilities is crucial for conclusively identifying non-Abelian behavior and could be achieved through shot-noise measurements at a single QPC. If the Abelian $\frac{1}{2}e$ quasiparticles are indeed responsible for the interference signal, developing techniques to facilitate the tunneling of fundamental quasiparticles at the QPC will be essential. Adjusting the saddle point potential or screening of the inter-edge interactions could affect the QPCs' characteristics in this way. Our observation of an apparent interfering charge being twice the fundamental one at $\nu=-\frac{2}{3}$ and $\nu=\frac{5}{3}$ indicates that Abelian hole-conjugate states can provide valuable insights on how to control tunneling of different quasiparticle types. Resolving this question in bilayer graphene could permit direct observation of non-Abelian statistics at a number of distinct FQH phases.

\pagebreak
\newpage
\section*{References}

\begin{thebibliography}{10}
\expandafter\ifx\csname url\endcsname\relax
  \def\url#1{\texttt{#1}}\fi
\expandafter\ifx\csname urlprefix\endcsname\relax\def\urlprefix{URL }\fi
\providecommand{\bibinfo}[2]{#2}
\providecommand{\eprint}[2][]{\url{#2}}

\bibitem{feldman2021fractional}
\bibinfo{author}{Feldman, D.~E.} \& \bibinfo{author}{Halperin, B.~I.}
\newblock \bibinfo{title}{Fractional charge and fractional statistics in the quantum {Hall} effects}.
\newblock \emph{\bibinfo{journal}{Reports on Progress in Physics}} \textbf{\bibinfo{volume}{84}}, \bibinfo{pages}{076501} (\bibinfo{year}{2021}).

\bibitem{nayak2008non}
\bibinfo{author}{Nayak, C.}, \bibinfo{author}{Simon, S.~H.}, \bibinfo{author}{Stern, A.}, \bibinfo{author}{Freedman, M.} \& \bibinfo{author}{Das~Sarma, S.}
\newblock \bibinfo{title}{Non-{A}belian anyons and topological quantum computation}.
\newblock \emph{\bibinfo{journal}{Reviews of Modern Physics}} \textbf{\bibinfo{volume}{80}}, \bibinfo{pages}{1083--1159} (\bibinfo{year}{2008}).

\bibitem{du2009fractional}
\bibinfo{author}{Du, X.}, \bibinfo{author}{Skachko, I.}, \bibinfo{author}{Duerr, F.}, \bibinfo{author}{Luican, A.} \& \bibinfo{author}{Andrei, E.~Y.}
\newblock \bibinfo{title}{Fractional quantum {Hall} effect and insulating phase of {D}irac electrons in graphene}.
\newblock \emph{\bibinfo{journal}{Nature}} \textbf{\bibinfo{volume}{462}}, \bibinfo{pages}{192--195} (\bibinfo{year}{2009}).

\bibitem{saminadayar1997observation}
\bibinfo{author}{Saminadayar, L.}, \bibinfo{author}{Glattli, D.}, \bibinfo{author}{Jin, Y.} \& \bibinfo{author}{Etienne, B. c.-m.}
\newblock \bibinfo{title}{Observation of the e/3 {F}ractionally {C}harged {L}aughlin {Q}uasiparticle}.
\newblock \emph{\bibinfo{journal}{Physical Review Letters}} \textbf{\bibinfo{volume}{79}}, \bibinfo{pages}{2526} (\bibinfo{year}{1997}).

\bibitem{de1998direct}
\bibinfo{author}{De-Picciotto, R.} \emph{et~al.}
\newblock \bibinfo{title}{Direct observation of a fractional charge}.
\newblock \emph{\bibinfo{journal}{Physica B: Condensed Matter}} \textbf{\bibinfo{volume}{249}}, \bibinfo{pages}{395--400} (\bibinfo{year}{1998}).

\bibitem{dolev2008observation}
\bibinfo{author}{Dolev, M.}, \bibinfo{author}{Heiblum, M.}, \bibinfo{author}{Umansky, V.}, \bibinfo{author}{Stern, A.} \& \bibinfo{author}{Mahalu, D.}
\newblock \bibinfo{title}{Observation of a quarter of an electron charge at the $\nu$ = 5/2 quantum {Hall} state}.
\newblock \emph{\bibinfo{journal}{Nature}} \textbf{\bibinfo{volume}{452}}, \bibinfo{pages}{829--834} (\bibinfo{year}{2008}).

\bibitem{venkatachalam2011local}
\bibinfo{author}{Venkatachalam, V.}, \bibinfo{author}{Yacoby, A.}, \bibinfo{author}{Pfeiffer, L.} \& \bibinfo{author}{West, K.}
\newblock \bibinfo{title}{Local charge of the $\nu$ = 5/2 fractional quantum {Hall} state}.
\newblock \emph{\bibinfo{journal}{Nature}} \textbf{\bibinfo{volume}{469}}, \bibinfo{pages}{185--188} (\bibinfo{year}{2011}).

\bibitem{MA2024324}
\bibinfo{author}{Ma, K.~K.}, \bibinfo{author}{Peterson, M.~R.}, \bibinfo{author}{Scarola, V.} \& \bibinfo{author}{Yang, K.}
\newblock \bibinfo{title}{Fractional quantum {Hall} effect at the filling factor $\nu$ = 5/2}.
\newblock \emph{\bibinfo{journal}{Encycl. Condens. Matter Phys.}} \textbf{\bibinfo{volume}{1}}, \bibinfo{pages}{324} (\bibinfo{year}{2024}).

\bibitem{halperin2011theory}
\bibinfo{author}{Halperin, B.~I.}, \bibinfo{author}{Stern, A.}, \bibinfo{author}{Neder, I.} \& \bibinfo{author}{Rosenow, B.}
\newblock \bibinfo{title}{Theory of the {F}abry-{P}{\'e}rot quantum {Hall} interferometer}.
\newblock \emph{\bibinfo{journal}{Physical Review B}} \textbf{\bibinfo{volume}{83}}, \bibinfo{pages}{155440} (\bibinfo{year}{2011}).

\bibitem{nakamura2019aharonov}
\bibinfo{author}{Nakamura, J.} \emph{et~al.}
\newblock \bibinfo{title}{Aharonov--{B}ohm interference of fractional quantum {Hall} edge modes}.
\newblock \emph{\bibinfo{journal}{Nature Physics}} \textbf{\bibinfo{volume}{15}}, \bibinfo{pages}{563--569} (\bibinfo{year}{2019}).

\bibitem{nakamura2020direct}
\bibinfo{author}{Nakamura, J.}, \bibinfo{author}{Liang, S.}, \bibinfo{author}{Gardner, G.~C.} \& \bibinfo{author}{Manfra, M.~J.}
\newblock \bibinfo{title}{Direct observation of anyonic braiding statistics}.
\newblock \emph{\bibinfo{journal}{Nature Physics}} \textbf{\bibinfo{volume}{16}}, \bibinfo{pages}{931--936} (\bibinfo{year}{2020}).

\bibitem{nakamura2023fabry}
\bibinfo{author}{Nakamura, J.}, \bibinfo{author}{Liang, S.}, \bibinfo{author}{Gardner, G.~C.} \& \bibinfo{author}{Manfra, M.~J.}
\newblock \bibinfo{title}{Fabry-{P}{\'e}rot {I}nterferometry at the $\nu$ = 2/5 {F}ractional {Q}uantum {Hall} {S}tate}.
\newblock \emph{\bibinfo{journal}{Physical Review X}} \textbf{\bibinfo{volume}{13}}, \bibinfo{pages}{041012} (\bibinfo{year}{2023}).

\bibitem{kim2024aharonov}
\bibinfo{author}{Kim, J.} \emph{et~al.}
\newblock \bibinfo{title}{Aharonov--{B}ohm interference and statistical phase-jump evolution in fractional quantum {Hall} states in bilayer graphene}.
\newblock \emph{\bibinfo{journal}{Nature Nanotechnology}} \bibinfo{pages}{1--8} (\bibinfo{year}{2024}).

\bibitem{werkmeister2024anyon}
\bibinfo{author}{Werkmeister, T.} \emph{et~al.}
\newblock \bibinfo{title}{Anyon braiding and telegraph noise in a graphene interferometer}.
\newblock \emph{\bibinfo{journal}{cond-mat/2403.18983}}  (\bibinfo{year}{2024}).

\bibitem{samuelson2024anyonic}
\bibinfo{author}{Samuelson, N.~L.} \emph{et~al.}
\newblock \bibinfo{title}{Anyonic statistics and slow quasiparticle dynamics in a graphene fractional quantum {Hall} interferometer}.
\newblock \emph{\bibinfo{journal}{cond-mat/2403.19628}}  (\bibinfo{year}{2024}).

\bibitem{ghosh2024anyonic}
\bibinfo{author}{Ghosh, B.} \emph{et~al.}
\newblock \bibinfo{title}{Anyonic {B}raiding in a {C}hiral {M}ach-{Z}ehnder {I}nterferometer}.
\newblock \emph{\bibinfo{journal}{cond-mat/2410.16488}}  (\bibinfo{year}{2024}).

\bibitem{bartolomei2020fractional}
\bibinfo{author}{Bartolomei, H.} \emph{et~al.}
\newblock \bibinfo{title}{Fractional statistics in anyon collisions}.
\newblock \emph{\bibinfo{journal}{Science}} \textbf{\bibinfo{volume}{368}}, \bibinfo{pages}{173--177} (\bibinfo{year}{2020}).

\bibitem{taktak2022two}
\bibinfo{author}{Taktak, I.} \emph{et~al.}
\newblock \bibinfo{title}{Two-particle time-domain interferometry in the fractional quantum {Hall} effect regime}.
\newblock \emph{\bibinfo{journal}{Nature Communications}} \textbf{\bibinfo{volume}{13}}, \bibinfo{pages}{5863} (\bibinfo{year}{2022}).

\bibitem{ruelle2023comparing}
\bibinfo{author}{Ruelle, M.} \emph{et~al.}
\newblock \bibinfo{title}{Comparing {F}ractional {Q}uantum {Hall} {L}aughlin and {J}ain {T}opological {O}rders with the {A}nyon {C}ollider}.
\newblock \emph{\bibinfo{journal}{Physical Review X}} \textbf{\bibinfo{volume}{13}}, \bibinfo{pages}{011031} (\bibinfo{year}{2023}).

\bibitem{lee2023partitioning}
\bibinfo{author}{Lee, J.-Y.~M.} \emph{et~al.}
\newblock \bibinfo{title}{Partitioning of diluted anyons reveals their braiding statistics}.
\newblock \emph{\bibinfo{journal}{Nature}} \textbf{\bibinfo{volume}{617}}, \bibinfo{pages}{277--281} (\bibinfo{year}{2023}).

\bibitem{glidic2023cross}
\bibinfo{author}{Glidic, P.} \emph{et~al.}
\newblock \bibinfo{title}{Cross-{C}orrelation {I}nvestigation of {A}nyon {S}tatistics in the $\nu$ = 1/3 and 2/5 {F}ractional {Q}uantum {Hall} {S}tates}.
\newblock \emph{\bibinfo{journal}{Physical Review X}} \textbf{\bibinfo{volume}{13}}, \bibinfo{pages}{011030} (\bibinfo{year}{2023}).

\bibitem{glidic2024signature}
\bibinfo{author}{Glidic, P.} \emph{et~al.}
\newblock \bibinfo{title}{Signature of anyonic statistics in the integer quantum {Hall} regime}.
\newblock \emph{\bibinfo{journal}{Nature Communications}} \textbf{\bibinfo{volume}{15}}, \bibinfo{pages}{6578} (\bibinfo{year}{2024}).

\bibitem{rosenow2016current}
\bibinfo{author}{Rosenow, B.}, \bibinfo{author}{Levkivskyi, I.~P.} \& \bibinfo{author}{Halperin, B.~I.}
\newblock \bibinfo{title}{Current {C}orrelations from a {M}esoscopic {A}nyon {C}ollider}.
\newblock \emph{\bibinfo{journal}{Physical Review Letters}} \textbf{\bibinfo{volume}{116}}, \bibinfo{pages}{156802} (\bibinfo{year}{2016}).

\bibitem{willett2023interference}
\bibinfo{author}{Willett, R.} \emph{et~al.}
\newblock \bibinfo{title}{Interference {M}easurements of {N}on-{Abelian} e/4 \& {Abelian} e/2 {Q}uasiparticle {B}raiding}.
\newblock \emph{\bibinfo{journal}{Physical Review X}} \textbf{\bibinfo{volume}{13}}, \bibinfo{pages}{011028} (\bibinfo{year}{2023}).

\bibitem{willett1987observation}
\bibinfo{author}{Willett, R.} \emph{et~al.}
\newblock \bibinfo{title}{Observation of an even-denominator quantum number in the fractional quantum {Hall} effect}.
\newblock \emph{\bibinfo{journal}{Physical Review Letters}} \textbf{\bibinfo{volume}{59}}, \bibinfo{pages}{1776} (\bibinfo{year}{1987}).

\bibitem{falson2015even}
\bibinfo{author}{Falson, J.} \emph{et~al.}
\newblock \bibinfo{title}{Even-denominator fractional quantum {Hall} physics in {Z}n{O}}.
\newblock \emph{\bibinfo{journal}{Nature Physics}} \textbf{\bibinfo{volume}{11}}, \bibinfo{pages}{347--351} (\bibinfo{year}{2015}).

\bibitem{narayanan2018incompressible}
\bibinfo{author}{Narayanan, S.}, \bibinfo{author}{Roy, B.} \& \bibinfo{author}{Kennett, M.~P.}
\newblock \bibinfo{title}{Incompressible even denominator fractional quantum {Hall} states in the zeroth {Landau} level of monolayer graphene}.
\newblock \emph{\bibinfo{journal}{Physical Review B}} \textbf{\bibinfo{volume}{98}}, \bibinfo{pages}{235411} (\bibinfo{year}{2018}).

\bibitem{kim2019even}
\bibinfo{author}{Kim, Y.} \emph{et~al.}
\newblock \bibinfo{title}{Even denominator fractional quantum {Hall} states in higher {Landau} levels of graphene}.
\newblock \emph{\bibinfo{journal}{Nature Physics}} \textbf{\bibinfo{volume}{15}}, \bibinfo{pages}{154--158} (\bibinfo{year}{2019}).

\bibitem{ki2014observation}
\bibinfo{author}{Ki, D.-K.}, \bibinfo{author}{Fal’ko, V.~I.}, \bibinfo{author}{Abanin, D.~A.} \& \bibinfo{author}{Morpurgo, A.~F.}
\newblock \bibinfo{title}{Observation of {E}ven {D}enominator {F}ractional {Q}uantum {Hall} {E}ffect in {S}uspended {B}ilayer {B}raphene}.
\newblock \emph{\bibinfo{journal}{Nano Letters}} \textbf{\bibinfo{volume}{14}}, \bibinfo{pages}{2135--2139} (\bibinfo{year}{2014}).

\bibitem{li2017even}
\bibinfo{author}{Li, J.} \emph{et~al.}
\newblock \bibinfo{title}{Even-denominator fractional quantum {Hall} states in bilayer graphene}.
\newblock \emph{\bibinfo{journal}{Science}} \textbf{\bibinfo{volume}{358}}, \bibinfo{pages}{648--652} (\bibinfo{year}{2017}).

\bibitem{zibrov2017tunable}
\bibinfo{author}{Zibrov, A.~A.} \emph{et~al.}
\newblock \bibinfo{title}{Tunable interacting composite fermion phases in a half-filled bilayer-graphene {Landau} level}.
\newblock \emph{\bibinfo{journal}{Nature}} \textbf{\bibinfo{volume}{549}}, \bibinfo{pages}{360--364} (\bibinfo{year}{2017}).

\bibitem{zibrov2018even}
\bibinfo{author}{Zibrov, A.} \emph{et~al.}
\newblock \bibinfo{title}{Even-denominator fractional quantum {Hall} states at an isospin transition in monolayer graphene}.
\newblock \emph{\bibinfo{journal}{Nature Physics}} \textbf{\bibinfo{volume}{14}}, \bibinfo{pages}{930--935} (\bibinfo{year}{2018}).

\bibitem{kumar2024quarter}
\bibinfo{author}{Kumar, R.} \emph{et~al.}
\newblock \bibinfo{title}{Quarter-and half-filled quantum {Hall} states and their competing interactions in bilayer graphene}.
\newblock \emph{\bibinfo{journal}{cond-mat/2405.19405}}  (\bibinfo{year}{2024}).

\bibitem{shi2020odd}
\bibinfo{author}{Shi, Q.} \emph{et~al.}
\newblock \bibinfo{title}{Odd-and even-denominator fractional quantum {Hall} states in monolayer wse2}.
\newblock \emph{\bibinfo{journal}{Nature Nanotechnology}} \textbf{\bibinfo{volume}{15}}, \bibinfo{pages}{569--573} (\bibinfo{year}{2020}).

\bibitem{banerjee2018observation}
\bibinfo{author}{Banerjee, M.} \emph{et~al.}
\newblock \bibinfo{title}{Observation of half-integer thermal {Hall} conductance}.
\newblock \emph{\bibinfo{journal}{Nature}} \textbf{\bibinfo{volume}{559}}, \bibinfo{pages}{205--210} (\bibinfo{year}{2018}).

\bibitem{dutta2022distinguishing}
\bibinfo{author}{Dutta, B.} \emph{et~al.}
\newblock \bibinfo{title}{Distinguishing between non-abelian topological orders in a quantum {Hall} system}.
\newblock \emph{\bibinfo{journal}{Science}} \textbf{\bibinfo{volume}{375}}, \bibinfo{pages}{193--197} (\bibinfo{year}{2022}).

\bibitem{son2015composite}
\bibinfo{author}{Son, D.~T.}
\newblock \bibinfo{title}{Is the {C}omposite {F}ermion a {Dirac} {P}article?}
\newblock \emph{\bibinfo{journal}{Physical Review X}} \textbf{\bibinfo{volume}{5}}, \bibinfo{pages}{031027} (\bibinfo{year}{2015}).

\bibitem{moore1991nonabelions}
\bibinfo{author}{Moore, G.} \& \bibinfo{author}{Read, N.}
\newblock \bibinfo{title}{Nonabelions in the fractional quantum {Hall} effect}.
\newblock \emph{\bibinfo{journal}{Nuclear Physics B}} \textbf{\bibinfo{volume}{360}}, \bibinfo{pages}{362--396} (\bibinfo{year}{1991}).

\bibitem{lee2007particle}
\bibinfo{author}{Lee, S.-S.}, \bibinfo{author}{Ryu, S.}, \bibinfo{author}{Nayak, C.} \& \bibinfo{author}{Fisher, M.~P.}
\newblock \bibinfo{title}{Particle-{H}ole {S}ymmetry and the $\nu$ = 5/2 {Q}uantum {Hall} {S}tate}.
\newblock \emph{\bibinfo{journal}{Physical Review Letters}} \textbf{\bibinfo{volume}{99}}, \bibinfo{pages}{236807} (\bibinfo{year}{2007}).

\bibitem{levin2007particle}
\bibinfo{author}{Levin, M.}, \bibinfo{author}{Halperin, B.~I.} \& \bibinfo{author}{Rosenow, B.}
\newblock \bibinfo{title}{Particle-{H}ole {S}ymmetry and the {Pfaffian} {S}tate}.
\newblock \emph{\bibinfo{journal}{Physical Review Letters}} \textbf{\bibinfo{volume}{99}}, \bibinfo{pages}{236806} (\bibinfo{year}{2007}).

\bibitem{levin2009collective}
\bibinfo{author}{Levin, M.} \& \bibinfo{author}{Halperin, B.~I.}
\newblock \bibinfo{title}{Collective states of non-{Abelian} quasiparticles in a magnetic field}.
\newblock \emph{\bibinfo{journal}{Physical Review B}} \textbf{\bibinfo{volume}{79}}, \bibinfo{pages}{205301} (\bibinfo{year}{2009}).

\bibitem{yutushui2024paired}
\bibinfo{author}{Yutushui, M.}, \bibinfo{author}{Hermanns, M.} \& \bibinfo{author}{Mross, D.~F.}
\newblock \bibinfo{title}{Paired fermions in strong magnetic fields and daughters of even-denominator {Hall} plateaus}.
\newblock \emph{\bibinfo{journal}{Physical Review B}} \textbf{\bibinfo{volume}{110}}, \bibinfo{pages}{165402} (\bibinfo{year}{2024}).

\bibitem{zheltonozhskii2024identifying}
\bibinfo{author}{Zheltonozhskii, E.}, \bibinfo{author}{Stern, A.} \& \bibinfo{author}{Lindner, N.}
\newblock \bibinfo{title}{Identifying the topological order of quantized half-filled {Landau} levels through their daughter states}.
\newblock \emph{\bibinfo{journal}{cond-mat/2405.03780}}  (\bibinfo{year}{2024}).

\bibitem{zhang2024hierarchy}
\bibinfo{author}{Zhang, C.}, \bibinfo{author}{Vishwanath, A.} \& \bibinfo{author}{Wen, X.-G.}
\newblock \bibinfo{title}{Hierarchy construction for non-abelian fractional quantum {Hall} states via anyon condensation}.
\newblock \emph{\bibinfo{journal}{cond-mat/2406.12068}}  (\bibinfo{year}{2024}).

\bibitem{bid2009shot}
\bibinfo{author}{Bid, A.}, \bibinfo{author}{Ofek, N.}, \bibinfo{author}{Heiblum, M.}, \bibinfo{author}{Umansky, V.} \& \bibinfo{author}{Mahalu, D.}
\newblock \bibinfo{title}{Shot {N}oise and {C}harge at the 2/3 {C}omposite {F}ractional {Q}uantum {Hall} {S}tate}.
\newblock \emph{\bibinfo{journal}{Physical Review Letters}} \textbf{\bibinfo{volume}{103}}, \bibinfo{pages}{236802} (\bibinfo{year}{2009}).

\bibitem{bhattacharyya2019melting}
\bibinfo{author}{Bhattacharyya, R.}, \bibinfo{author}{Banerjee, M.}, \bibinfo{author}{Heiblum, M.}, \bibinfo{author}{Mahalu, D.} \& \bibinfo{author}{Umansky, V.}
\newblock \bibinfo{title}{Melting of {I}nterference in the {F}ractional {Q}uantum {Hall} {E}ffect: Appearance of {N}eutral {M}odes}.
\newblock \emph{\bibinfo{journal}{Physical Review Letters}} \textbf{\bibinfo{volume}{122}}, \bibinfo{pages}{246801} (\bibinfo{year}{2019}).

\bibitem{arovas1984fractional}
\bibinfo{author}{Arovas, D.}, \bibinfo{author}{Schrieffer, J.~R.} \& \bibinfo{author}{Wilczek, F.}
\newblock \bibinfo{title}{Fractional {S}tatistics and the {Q}uantum {Hall} {E}ffect}.
\newblock \emph{\bibinfo{journal}{Physical Review Letters}} \textbf{\bibinfo{volume}{53}}, \bibinfo{pages}{722} (\bibinfo{year}{1984}).

\bibitem{kivelson1990semiclassical}
\bibinfo{author}{Kivelson, S.}
\newblock \bibinfo{title}{Semiclassical theory of localized many-anyon states}.
\newblock \emph{\bibinfo{journal}{Physical Review Letters}} \textbf{\bibinfo{volume}{65}}, \bibinfo{pages}{3369} (\bibinfo{year}{1990}).

\bibitem{chamon1997two}
\bibinfo{author}{Chamon, C. d.~C.}, \bibinfo{author}{Freed, D.}, \bibinfo{author}{Kivelson, S.}, \bibinfo{author}{Sondhi, S.} \& \bibinfo{author}{Wen, X.}
\newblock \bibinfo{title}{Two point-contact interferometer for quantum {Hall} systems}.
\newblock \emph{\bibinfo{journal}{Physical Review B}} \textbf{\bibinfo{volume}{55}}, \bibinfo{pages}{2331} (\bibinfo{year}{1997}).

\bibitem{stern2006proposed}
\bibinfo{author}{Stern, A.} \& \bibinfo{author}{Halperin, B.~I.}
\newblock \bibinfo{title}{Proposed {E}xperiments to {P}robe the non-{Abelian} $\nu$ = 5/2 {Q}uantum {Hall} {S}tate}.
\newblock \emph{\bibinfo{journal}{Physical Review Letters}} \textbf{\bibinfo{volume}{96}}, \bibinfo{pages}{016802} (\bibinfo{year}{2006}).

\bibitem{bonderson2006detecting}
\bibinfo{author}{Bonderson, P.}, \bibinfo{author}{Kitaev, A.} \& \bibinfo{author}{Shtengel, K.}
\newblock \bibinfo{title}{Detecting {N}on-{Abelian} {S}tatistics in the $\nu$ = 5/2 {F}ractional {Q}uantum {Hall} {S}tate}.
\newblock \emph{\bibinfo{journal}{Physical Review Letters}} \textbf{\bibinfo{volume}{96}}, \bibinfo{pages}{016803} (\bibinfo{year}{2006}).

\bibitem{huang2022valley}
\bibinfo{author}{Huang, K.} \emph{et~al.}
\newblock \bibinfo{title}{Valley {I}sospin {C}ontrolled {F}ractional {Q}uantum {Hall} {S}tates in {B}ilayer {G}raphene}.
\newblock \emph{\bibinfo{journal}{Physical Review X}} \textbf{\bibinfo{volume}{12}}, \bibinfo{pages}{031019} (\bibinfo{year}{2022}).

\bibitem{hu2023high}
\bibinfo{author}{Hu, Y.} \emph{et~al.}
\newblock \bibinfo{title}{High-{R}esolution {T}unneling {T}pectroscopy of {F}ractional {Q}uantum {Hall} {S}tates}.
\newblock \emph{\bibinfo{journal}{cond-mat/2308.05789}}  (\bibinfo{year}{2023}).

\bibitem{assouline2024energy}
\bibinfo{author}{Assouline, A.} \emph{et~al.}
\newblock \bibinfo{title}{Energy {G}ap of the {E}ven-denominator {F}ractional {Q}uantum {Hall} {S}tate in {B}ilayer {G}raphene}.
\newblock \emph{\bibinfo{journal}{Physical Review Letters}} \textbf{\bibinfo{volume}{132}}, \bibinfo{pages}{046603} (\bibinfo{year}{2024}).

\bibitem{ghosh2024bunching}
\bibinfo{author}{Ghosh, B.}, \bibinfo{author}{Labendik, M.}, \bibinfo{author}{Umansky, V.}, \bibinfo{author}{Heiblum, M.} \& \bibinfo{author}{Mross, D.~F.}
\newblock \bibinfo{title}{Coherent {B}unching of {A}nyons and their {D}issociation in {I}nterference {E}xperiments}.
\newblock \emph{\bibinfo{journal}{cond-mat/2412.16316}}  (\bibinfo{year}{2024}).

\bibitem{kane1994randomness}
\bibinfo{author}{Kane, C.}, \bibinfo{author}{Fisher, M.~P.} \& \bibinfo{author}{Polchinski, J.}
\newblock \bibinfo{title}{Randomness at the edge: Theory of quantum {Hall} transport at filling $\nu$ = 2/3}.
\newblock \emph{\bibinfo{journal}{Physical Review Letters}} \textbf{\bibinfo{volume}{72}}, \bibinfo{pages}{4129} (\bibinfo{year}{1994}).

\bibitem{chang2003chiral}
\bibinfo{author}{Chang, A.}
\newblock \bibinfo{title}{Chiral {Luttinger} liquids at the fractional quantum {Hall} edge}.
\newblock \emph{\bibinfo{journal}{Reviews of Modern Physics}} \textbf{\bibinfo{volume}{75}}, \bibinfo{pages}{1449} (\bibinfo{year}{2003}).

\bibitem{radu2008quasi}
\bibinfo{author}{Radu, I.~P.} \emph{et~al.}
\newblock \bibinfo{title}{Quasi-{P}article {P}roperties from {T}unneling in the $\nu$ = 5/2 {F}ractional {Q}uantum {Hall} {S}tate}.
\newblock \emph{\bibinfo{journal}{Science}} \textbf{\bibinfo{volume}{320}}, \bibinfo{pages}{899--902} (\bibinfo{year}{2008}).

\bibitem{lin2012measurements}
\bibinfo{author}{Lin, X.}, \bibinfo{author}{Dillard, C.}, \bibinfo{author}{Kastner, M.}, \bibinfo{author}{Pfeiffer, L.} \& \bibinfo{author}{West, K.}
\newblock \bibinfo{title}{Measurements of quasiparticle tunneling in the $\nu$ = 5/2 fractional quantum {H}all state}.
\newblock \emph{\bibinfo{journal}{Physical Review B}} \textbf{\bibinfo{volume}{85}}, \bibinfo{pages}{165321} (\bibinfo{year}{2012}).

\end{thebibliography}

\pagebreak
\section*{Materials and Methods}
\subsection{Stack Preparation $\colon$}
In this study, we employ van der Waals (vdW) heterostructures in which a bilayer graphene layer is encapsulated between hexagonal boron nitride (hBN) and graphite layers. To prepare flakes from bulk graphite crystals, SiO$_{2}$/Si substrates are cut into 10 mm $\times$10 mm pieces. These pieces are placed on tape, which is used to exfoliate the bulk crystals. The exfoliated flakes are then transferred onto the SiO$_{2}$/Si substrate and heated on a hot plate at 170–180°C for 90 seconds. When the tape has cooled, these pieces are removed to search for the desired bilayer graphene and graphite flakes. After cooling, the tape is removed, and the flakes are inspected to identify suitable bilayer graphene and graphite layers. hBN flakes are prepared in a similar manner by exfoliating bulk hBN crystals using thin polydimethylsiloxane (PDMS). The stack is prepared using polycarbonate stamps held with Kapton tape, placed on a diamond-shaped PDMS layer atop a glass slide. To ensure strong adhesion between the polycarbonate film and the PDMS, the stamps are placed on a hot plate at 170–180°C for 2 hours. The transfer stage is heated to 130-131°C, allowing the sequential pickup of all vdW layers in the order: top graphite, top hBN, bilayer graphene, bottom hBN, and bottom graphite. A thickness of 29 (27) nm for the top (bottom) hBN and 5 nm for the top and bottom graphite is used in the fabricated device. The prepared stack is transferred at 180°C on a clean SiO$_{2}$/Si substrate and left for roughly 15 min at 180°C to melt the polycarbonate and detach it from the PDMS. The polycarbonate film is dissolved by placing the sample for 3–4 h in chloroform and subsequently cleaned with isopropyl alcohol (IPA) and deionized water. Subsequently, the stack on the SiO$_{2}$/Si substrate undergoes thermal annealing in an ultrahigh vacuum ($\sim $ 10$^{-9}$ torr) at 400°C for 4 hours and 30 mins to remove residual contaminants and bubbles. Finally, atomic force microscopy (AFM) ironing is performed to clean and flatten the local area where the device will be fabricated.

\subsection{Device fabrication $\colon$}
The bilayer graphene-based electronic Fabry-Perot interferometer (FPI) is fabricated on a five-layer vdW heterostructure placed on a highly p-doped Si substrate with a 280 nm SiO$_{2}$ oxide layer, using standard nanofabrication and lithography techniques. The process begins with creating alignment markers for the electron beam lithography and bonding pads, using Ti (10 nm)/Au (60 nm)/Pd (20 nm). The device geometry is defined through reactive ion etching (RIE) with polymethyl methacrylate (PMMA) resist serving as the etch mask. To etch out two main materials used in the heterostructure, hBN, and few-layer graphite layers, O$_{2}$/CHF$_{3}$ mixture with the volume ratio of 1:10, and O$_{2}$ are used as etching gas for hBN and graphite, respectively. After defining the geometry, the sample undergoes thermal annealing in ultrahigh vacuum ($\sim$ 10 $^{-9}$ Torr) at 350\textdegree{C} for 2 hours and 30 minutes for resist residual removal on the stack. The edge contacts are fabricated in the final step by etching the top hBN layer with O$_{2}$/CHF$_{3}$, followed by angled evaporation of Cr (2 nm)/Pd (20 nm)/Au (60 nm). A trench approximately 40 nm wide is etched into the top graphite using mild O$_{2}$ plasma conditions to minimize damage to the top hBN layer, dividing the top graphite into eight sections. Finally, bridges are fabricated to independently tune the potential of each graphite section. This is achieved using PMMA/MMA/PMMA trilayer resists, followed by a 20 seconds mild O$_{2}$ plasma etch and the subsequent evaporation of Cr (5 nm)/Au (320 nm).

\subsection{Measurements$\colon$}
The device is measured in a highly filtered dilution refrigerator at the base temperature of 10 mK utilizing the standard low-frequency lock-in amplifier technique. An SRS 865A lock-in amplifier generates an alternating voltage at 13.7 Hz and measures the voltage difference between two contacts. A 100 M$\Omega$ load resistor is included in series with the lock-in amplifier, allowing the system to source an alternating current ranging from 50 nA to 0.5 nA. A QDAC, an ultralow-noise 24-channel digital-to-analog converter (Qdevil-QM), is used to tune the voltages applied to all graphite gates and the two air bridges. Additionally, a Keithley 2400 voltage source is used to apply a voltage to the highly p-doped Si substrate, doping the contact region and improving the contact resistance.

\section*{Data and materials availability:}
The data supporting the plots in this paper and other findings of this study are available from the corresponding author upon request.

 \section*{Acknowledgements}
It is a pleasure to thank Moty Heiblum, Bert Halperin, and Philip Kim for illuminating discussions. \textbf{Funding:} J.K. acknowledges support from the Dean of the Faculty and the Clore Foundation. Y.R. acknowledges the support from the Quantum Science and Technology Program 2021, the Schwartz Reisman Collaborative Science Program, supported by the Gerald Schwartz and Heather Reisman Foundation, supported by a research grant from the Goldfield Family Charitable Trust, the Minerva Foundation with funding from the Federal German Ministry for Education and Research, and the European Research Council Starting Investigator Grant Anyons, 101163917. D.F.M acknowledges support from the ISF (grant 2572/21), from the Minerva Foundation with funding from the Federal German Ministry for Education and Research, and the DFG (CRC/Transregio 183). A.S. acknowledges support from the ISF, ISF Quantum Science and Technology (grant 2074/19), and the DFG (CRC/Transregio 183).

\section*{Author contributions}
H.D. prepared the stacks. K.W. and T.T. provided the hBN crystals. J.K., H.D., A.I., and R.K. improved the quality of the device. J.K. fabricated the device. A.H. and R.K. developed the measurement circuit and a dilution refrigerator. J.K. performed the measurements. J.K., H.D., A.S., A.S., D.M, and Y.R. analyzed the measured data. A.S., A.S., and D.M., developed the theoretical aspect. J.K., H.D., A.S., A.S., D.M., and Y.R. wrote the paper with input from all authors. Y.R. supervised the overall work done on the project.

\section*{Competing interests}
The authors declare no competing interests.
\newpage


\section*{Supplementary material}

# Supplementary Information for
# Aharonov–Bohm Interference in Even-Denominator Fractional Quantum Hall States

# Contents

## SI1  Device characterization

Fig. S1a shows the optical microscope image of the fabricated device for the current studies and Fig. S1b its scanning electron microscopic image. The scale bar is 200 nm. The red dashed hexagon shows the lithographic area of the Fabry-Pérot interference loop, which is $A = 1.0$ $\mu m^2$. To characterize the quality of the device, we measured the carrier mobility of the fabricated device in a Hall bar geometry. Fig. S1c shows the two-probe resistance $R_{2p}$, incorporating the contact resistance $R_c$ as a function of $V_{BG}$ while applying 5 nA at 0.1 K and $B = 0$ T. To extract the mobility $\mu$ we fit the measured $R_{2P}$ to

$$R_{2p} = R_c + \frac{L}{W} \cdot \frac{1}{\mu\sqrt{e^2 n_0^2 + C_{BG}^2(V_{BG} - V_{CNP})^2}}. \tag{S1}$$

Here, $n_0$ denotes the intrinsic carrier concentration induced by substrate doping, $C_{BG}$ is the capacitance per unit area of the bottom graphite gate, $V_{CNP}$ is the voltage corresponding to the charge neutrality point, and $L$ and $W$ are the length between the two probe contacts and width of the Hall bar channel, and $e$ the electron charge.[1]

By fitting to Eq. (S1), we obtained a high electron-mobility of $\mu \approx 1.23 \times 10^6$ cm$^{-2}$. Additionally, the intrinsic carrier concentration was found to be $n_0 = 1.4 \times 10^{-8}$ cm$^{-2}$, indicating a low charge carrier impurity density in our FPI device. In Fig. S1d, to confirm that the active layer consists of bilayer graphene, we observed its gap opening as a function of displacement field by measuring the two-probe resistance as a function of top gate voltage ($V_{TG}$) and bottom gate voltage ($V_{BG}$). As we move away from the charge neutrality point, the observed increase in resistance demonstrates the band gap opening in the bilayer graphene.

Following the low-field measurement to characterize the sample quality, we move on to characterize the device at high magnetic fields. Under the application of a magnetic field, 2D electrons in the bilayer are described by discrete Landau labels. When kinetic energy is quenched, electron interaction becomes crucial and gives rise to many incompressible states, such as the Fractional quantum Hall effect. We measure $R_{xx}$ as a function of top gate voltage ($V_{TG}$) and magnetic field ($B$) at a fixed back gate voltage ($V_{TG}$) of 0.65 V at $B$ = 11 T and 10 mK. We observe a rich fan diagram showing a series of even and odd fractional quantum Hall (FQH) states on both electron and hole sides, as shown in Fig. S1e,f, and consistent with earlier studies.[2,3] We observe between $\nu = 1, 2$ and between $\nu = -1, 0$, a half-filled plateau and few Jain states (near $D = 0$) while between $\nu = 0, 1$ and between $\nu = -2, 1$ there is no plateau at half-filled and many Jain states shown in Fig. S1e,f.

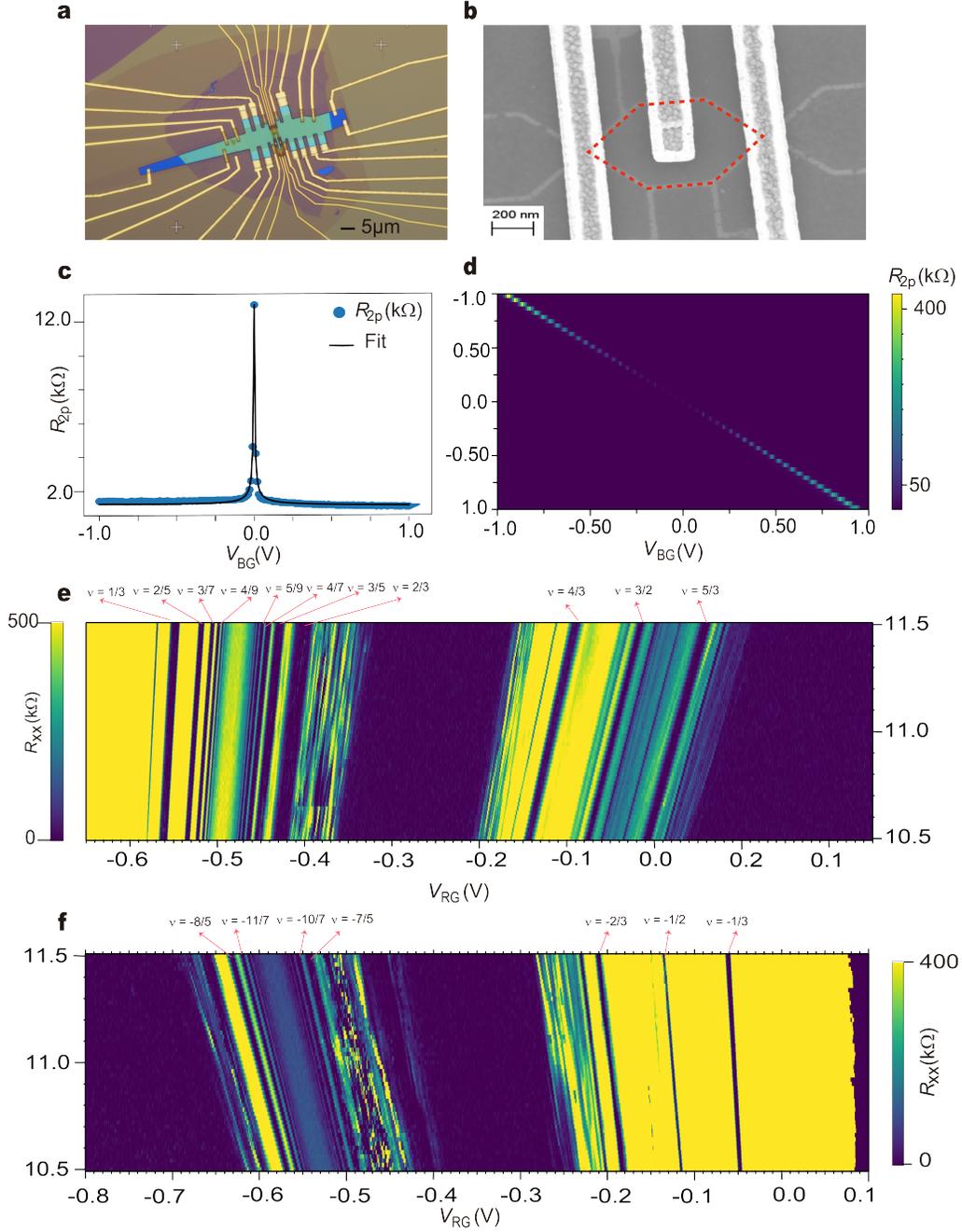

**Figure S1: Device Characterization** (**a**) Optical image of a bilayer graphene-based FPI. Scale bar 5$\mu$m (**b**) Scanning electron microscope image of the device, where the red dashed line shows an interference loop area of 1 $\mu m^2$. Scale bar 200nm (**c**) Two-probe resistance $R_{2p}$ as a function of $V_{BG}$, measured by applying a 5 nA current at 1 K under a magnetic field of 0 T. (**d**) $R_{2p}$ as a function of top and bottom gate voltage showing a clear gap opening in bilayer graphene, measured at 1 K at 0 T magnetic field. (**e**) $R_{xx}$ is measured at 10 mK and at 11 T with a fixed Si = 5 V and $V_{BG}$ = 0.65 V with 1 nA current applied. (**f**) $R_{xx}$ is measured at 10 mK and at 11 T with a fixed Si = -5 V and $V_{BG}$ = -0.1 V with 1 nA current applied.

2

## SI2 Extracting $\alpha_c$ from longitudinal resistance measurement

In this section, we briefly discuss the extraction of $\alpha_c$ values for different even and odd denominator fractional quantum Hall states. A 2D map of $R_{xx}$ is measured as a function of $V_{RG}$ at the base temperature of 10 mK between 10.5 T and 11.5 T as shown in Fig. 2b,e in the main text. To find the value of $\alpha_c$'s, from the 2D map, we take line cuts at two different magnetic fields, 10.5 T and 11.5 T. As shown in Fig. 2b,e, within the boundaries marked by the red dashed lines, we extract the slope of the constant filling line at the center of the $R_{xx}$ dip, which provides the values of $\alpha_c$'s.

**Determination of $\alpha_c$ for hole and electron carriers:** In Fig. S2a,b, we present fully developed FQH states measured at magnetic fields of 10.5 T and 11.5 T on the hole-doped and electron-doped sides, respectively.

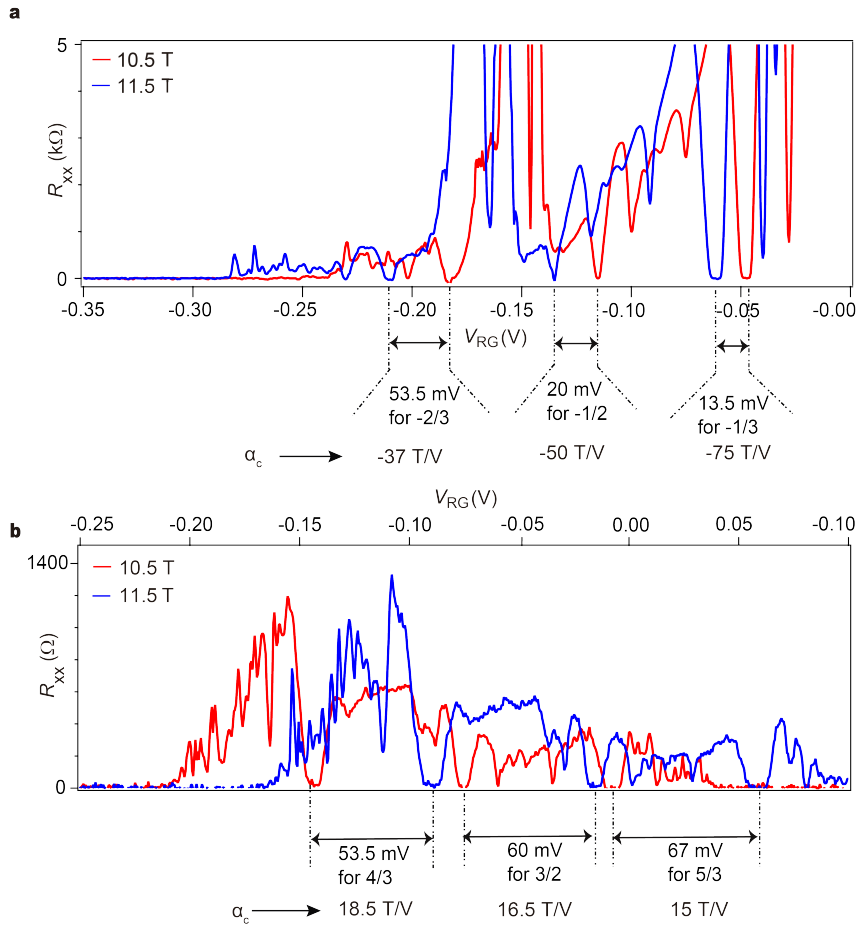

**Figure S2: Determination of $\alpha_C$ values.** (a) Line cuts of $R_{xx}$ as a function of $V_{RG}$ at magnetic fields of 10.5 T (red) and 11.5 T (blue) for hole carriers. (b) Same as (a), but for electron carriers.

We determine the shift in gate voltage, $\delta V_{RG}$, by calculating the difference between the positions of

3

two vertical dashed lines drawn at the center of the plateaus where $R_{\text{xx}}$ drops to zero for FQH states. For hole carriers, this corresponds to $\nu = -\frac{2}{3}, -\frac{1}{2}$, and $-\frac{1}{3}$ (Fig. S2a) and , and for electron carriers, $\nu = \frac{4}{3}, \frac{3}{2}$, and $\frac{5}{3}$ (Fig. S2b). The magnetic field difference, $\delta B$, is defined as the separation between the two magnetic field values at which the longitudinal resistance line cuts are taken. We then calculate $\alpha_c = \frac{\delta B}{\delta V_{\text{RG}}}$. For hole carriers, the FQHE states at $\nu = -\frac{2}{3}, -\frac{1}{2}$, and $-\frac{1}{3}$ yield $\alpha_c$ values of, -37 T/V, -50 T/V, and -75 T/V, respectively. For electron carriers, the FQHE states at $\nu = \frac{4}{3}, \frac{3}{2}$, and $\frac{5}{3}$ results in $\alpha_c$ values of 18.5 T/V, 16.5 T/V, and 15 T/V, respectively. From the Streda formula, $C_{\text{RG}} = \frac{e\alpha_c \nu}{\Phi_0}$, the capacitance of the right gate (RG) per unit area is calculated as $C_{\text{RG}} = 0.959$ mF/m$^2$ using the average value of $\alpha_c \nu$ for all fractional fillings, different from the value of $C_{\text{CG}}$ discussed in Sec. SI12.

## SI3 Electron carrier Aharonov-Bohm oscillations

In this section, we show diagonal resistance, $R_D$, as a function of plunger gate voltage, $V_{\text{PG}}$, measured at a constant magnetic field and a base temperature of 10 mK, with a source-drain current of 500 pA. This analysis is performed for both integer and fractional quantum Hall states on the electron-doped side. The quantum point contact (QPC) transmissions are maintained at approximately 50% for integers and 70–80% for fractional edge modes.

We set $\nu = 1, \frac{3}{2}, \frac{4}{3}$, and $\frac{5}{3}$ in the LG, CG, and RG regions and observe stable resistance oscillations as a function of the plunger gate voltage, $V_{\text{PG}}$. To estimate the visibility of the inner or outer edge modes, we use the following equation

$$\text{Visibility} = \frac{G_{\max} - G_{\min}}{G_{\max} + G_{\min} - 2 \cdot G_{\nu_{\text{outer}}}}, \tag{S2}$$

where $G_{\max}$ and $G_{\min}$, are the maximum and minimum diagonal conductance values, and $G_{\nu_{\text{outer}}}$ represents the conductance of a fully transmitted outer edge mode. The visibilities of the oscillations for $\nu = 1, \frac{3}{2}, \frac{4}{3}, \frac{5}{3}$ are roughly estimated to be 9.5%, 5.6%, 11.8%, 4.5% as shown in Fig. S3a-d. Notably, the hole-conjugate state at $\nu = \frac{5}{3}$ exhibits significantly lower visibility compared to the $\nu = \frac{4}{3}$ fractional quantum Hall state. One possible explanation for this difference is the distribution of current among multiple edge modes, with potentially only one mode contributing to interference. However, the primary reason for the lower visibility at $\nu = \frac{5}{3}$ compared to $\nu = \frac{4}{3}$ remains unclear.[4]

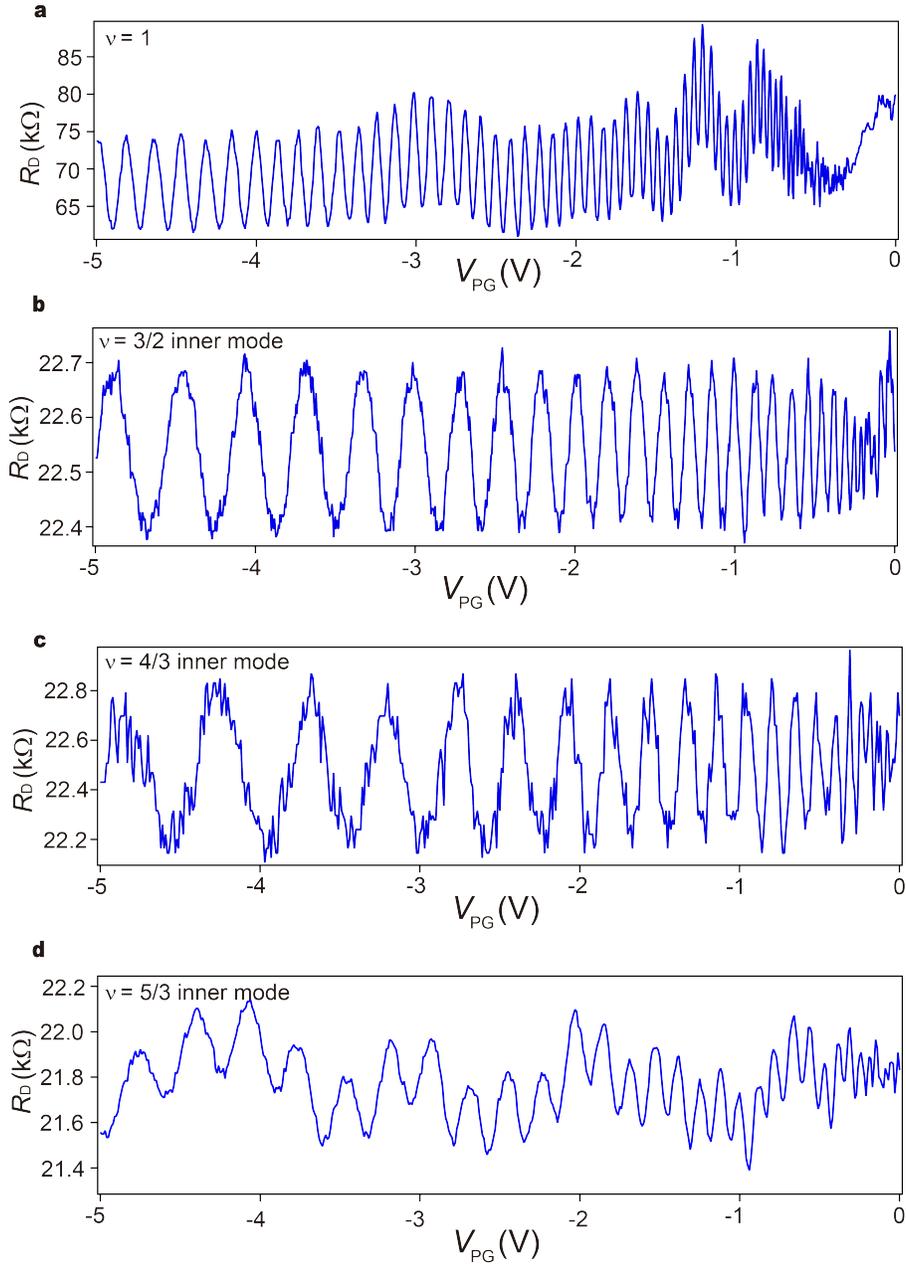

**Figure S3: Aharonov-Bohm oscillations as a function plunger gate voltage:** (**a-d**) Diagonal resistance measured as a function of plunger gate voltage for electron carriers for (a) $\nu = 1$, (b) $\nu = \frac{3}{2}$ inner mode, (c) $\nu = \frac{4}{3}$ inner mode, and (d) $\nu = \frac{5}{3}$ inner mode.

## SI4   Hole carrier Aharonov-Bohm oscillations

For hole carriers, we measure the diagonal resistance, $R_\text{D}$, as a function of plunger gate voltage, $V_\text{PG}$, at a constant magnetic field and a temperature of 10 mK, with a source-drain current of 100 pA. Figure S4a–e presents conductance oscillations for $\nu = -1$, $-\frac{1}{2}$, $-\frac{1}{3}$, $-\frac{2}{3}$, and $-2$, yielding visibilities of 2.3%, 1.9%, 2.3%, 0.7%, and 33.5%, respectively, using the formula defined in SI3. Similar to the electron-doped region, the diagonal resistance oscillations exhibit lower visibility for the particle-conjugate state $\nu = -\frac{2}{3}$ compared to the $\nu = -\frac{1}{3}$ fractional quantum Hall state.

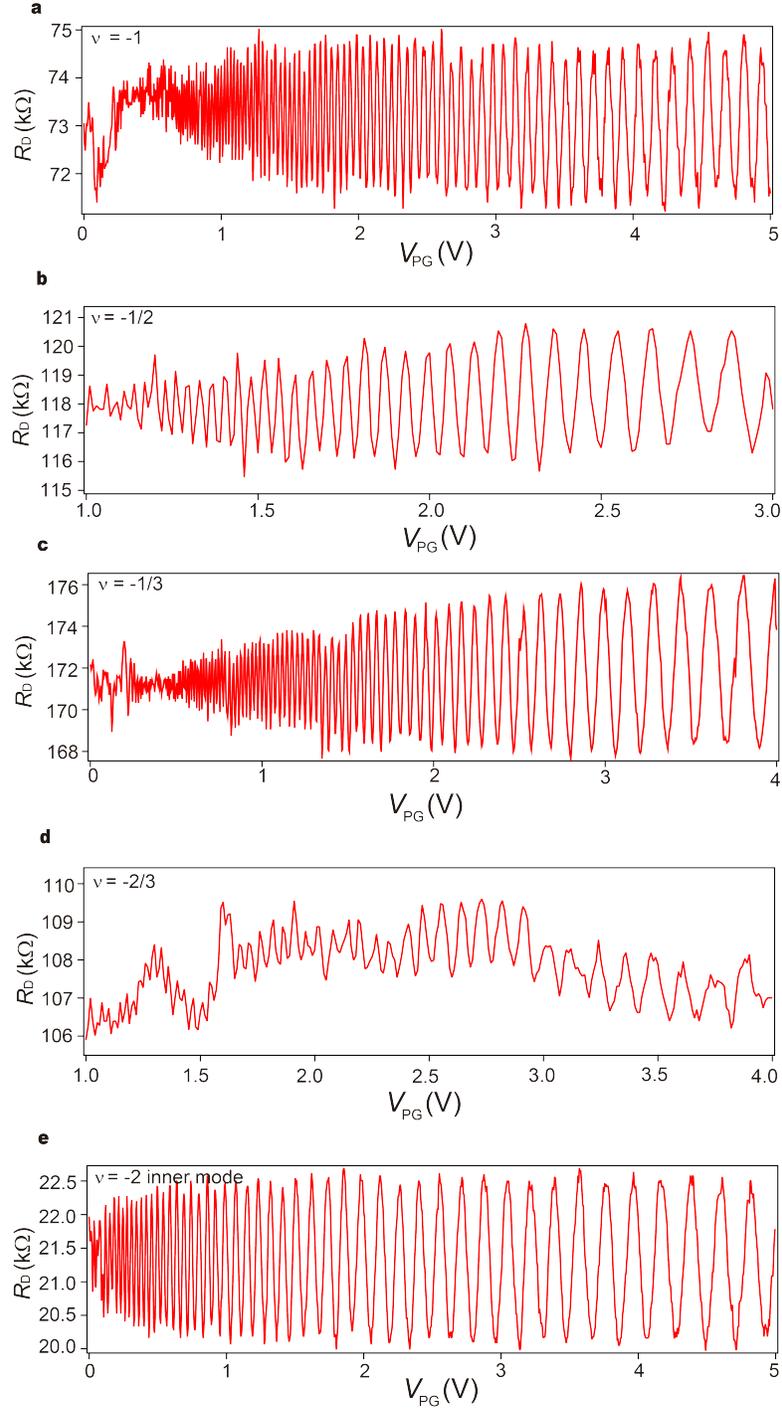

**Figure S4: Aharonov-Bohm oscillations as a function plunger gate voltage** (**a-e**) Diagonal resistance measured as a function of plunger gate voltage for the hole carriers for (a) $\nu = -1$, (b) $\nu = -\frac{1}{2}$, (c) $\nu = -\frac{1}{3}$, (d) $\nu = -\frac{2}{3}$, (e) $\nu = -2$ inner mode.

7

# SI5   1D-FFT analysis for the phase jump at $\nu = -\frac{1}{3}$

Fig. S5a shows diagonal resistance measurements as a function of the plunger gate voltage, $V_{\text{PG}}$, and the central gate voltage, $V_{\text{CG}}$, for $\alpha_c = 0$ at 11 T and 10 mK. In Fig. S5b, we focus on the region highlighted by the black dashed window, where a distinct series of phase slips is observed, corresponding to the addition or removal of charged quasiparticles. We perform 1D-FFTs of $\Delta R_{\text{D}}$ for fixed values of $V_{\text{CG}}$ to extract the magnitude of the phase jumps based on the phase at the 1D-FFT peak. To ensure a continuous phase evolution and avoid discontinuities at $\pm\pi$ as a function of $V_{\text{CG}}$, a $2\pi$ phase shift is added as necessary. The magnitude of the phase slips $\frac{\Delta\theta}{2\pi}$, are calculated by taking a difference between the average values on two adjacent AB regions after deducting the global AB contribution as shown in Fig. S5c. This analysis yields an average phase slip value of $\frac{\Delta\theta}{2\pi} = 0.30 \pm 0.03$, which is consistent with the theoretically expected value of $\theta_{\text{anyon}} = \frac{2\pi}{3}$.

The deviations in the values of the measured phase slips can arise from to bulk-edge coupling [5]

$$\frac{\Delta\theta}{2\pi} = -\frac{\theta_{\text{anyon}}}{2\pi} + 3\frac{K_{\text{IL}}}{K_{\text{I}}}\left(\frac{e^*}{e}\right)^2, \tag{S3}$$

where $K_{\text{IL}}$ is the bulk-to-edge coupling, and $K_{\text{I}}$ is the edge stiffness describing the energy cost to vary the interfering area $A$. According to Eq. (S3), the magnitude of the phase slips can vary when a non-negligible $K_{\text{IL}}$ exists, explaining that the majority of the observed phase slips are smaller than $\frac{2\pi}{3}$.[5,6] Moreover, the behavior of different $\frac{\Delta\theta}{2\pi}$ may be ascribed to the case in which both $K_{\text{IL}}$ and $K_{\text{IL}}$ can be affected by the number of bulk quasiparticles.

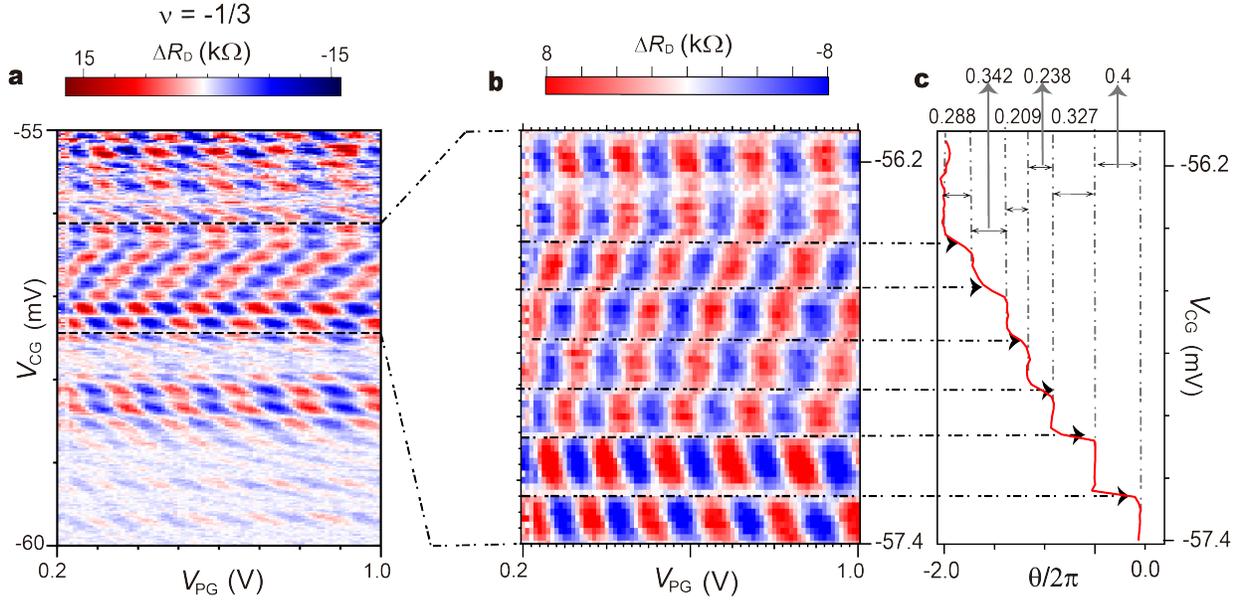

**Figure S5: 1D-FFT analysis for the phase jump for $\alpha = 0$** (**a**) $\Delta R_\text{D}$ is measured for $\nu = -\frac{1}{3}$ as function of $V_\text{CG}$ and $V_\text{PG}$ at 10 mK (**b**) A small window of $\Delta R_\text{D}$ from Fig. S5a where a series of phase slips occur. (**c**) $\frac{\theta}{2\pi}$ plotted as function of $V_\text{CG}$ with subtraction of the global Aharonov-Bohm contribution.

## SI6 Aharonov–Bohm interference in hole-doped odd- and even-denominator FQH states

Figs. S6a-d, show $R_\text{D}$ measurements plotted in $B|_{\alpha_c}$-$V_\text{PG}$ planes corresponding to filling factors of $\nu = -1$, $-\frac{2}{3}$, $-\frac{1}{2}$, and $-\frac{1}{3}$. For all fillings, clear AB-dominated interference patterns are observed. For hole doping, constant-phase lines in the AB regime are expected to exhibit positive slopes, in contrast to the negative slopes observed for electron doping. We observe a change in the $R_\text{D}$ background as a function of $V_\text{CG}$, which can be attributed to the change in the transmission of the QPCs. Figs. S6e-h present the 2D-FFTs of the corresponding data in Figs. S6a-d, respectively, plotted as a function of $\frac{\Phi_0}{\Delta B}$ and $\frac{1}{\Delta V_\text{PG}}$. We extract the interference area of 1.02 $\mu$m$^2$ for $\nu = -1$, assuming an interfering charge of $e$. This is in good agreement with the lithographic area of the center gate (CG) estimated to be 1 $\mu$m$^2$.

To determine the magnetic field periodicity $\Delta B$, we analyze one-dimensional cuts of the 2D-FFT as a function of $\frac{\Phi_0}{\Delta B}$, taken at the peak value of $\Delta V_\text{PG}$. Figs. S6i-l show these FFT amplitude line curves as a function of $\frac{\Phi_0}{\Delta B}$. By fitting a Gaussian curve to the data, we extract the mean value $\frac{\Phi_0}{\Delta B_\text{fit}}$ and the variance $\sigma^2$, as shown in the lower-right sides of Figs. S6i-l.

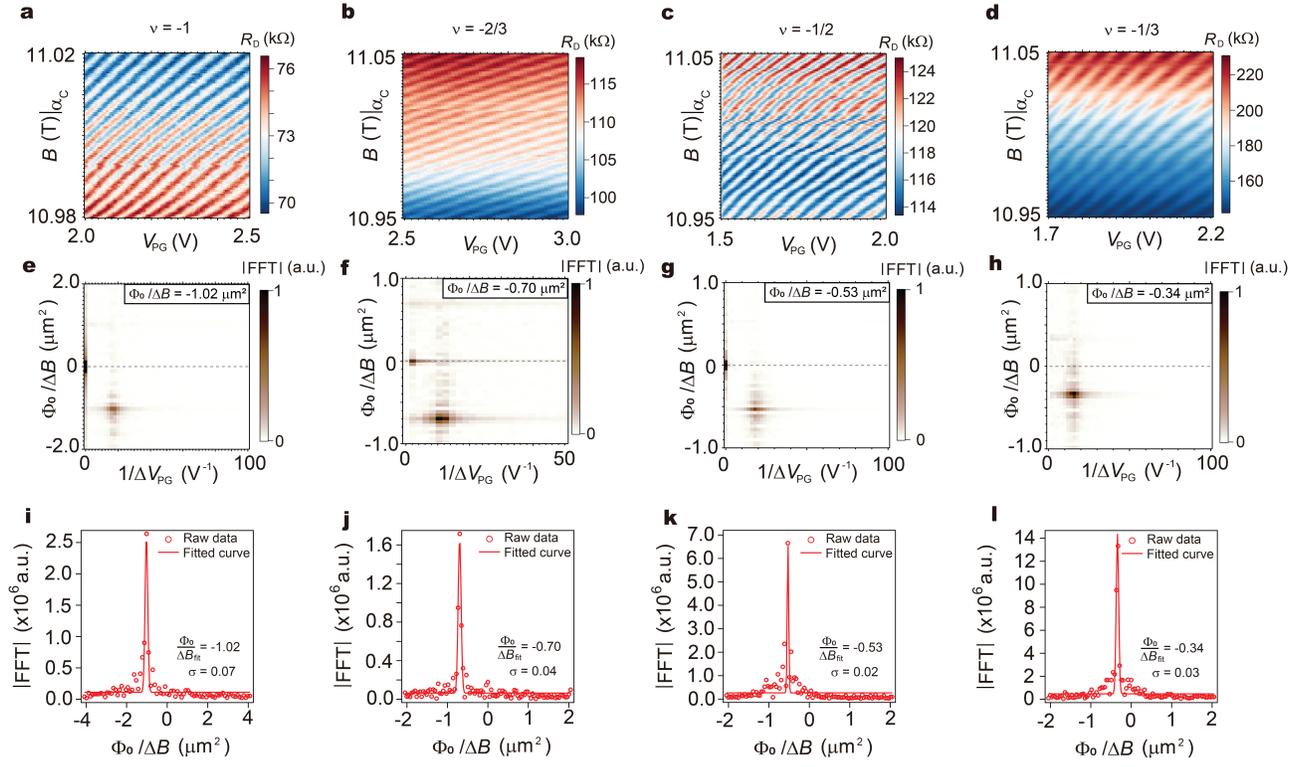

**Figure S6: Hole-doped Aharonov–Bohm interference.** (**a-d**) $R_D$ for $\nu = -1, -\frac{2}{3}, -\frac{1}{2}$, and $-\frac{1}{3}$, measured in $B|_{\alpha_c}$-$V_{PG}$ plane. (**e-h**) 2D-FFT as a function of $\frac{\Phi_0}{\Delta B}$ and $\frac{1}{\Delta V_{PG}}$, corresponding to Figs. S6a-d, respectively. (**i-l**) 1D cuts of the 2D-FFT as a function of $\frac{\Phi_0}{\Delta B}$, taken at the peak value of $\Delta V_{PG}$, obtained from Figs. S6e-h, respectively. Lower-right sides: Gaussian fitting parameters.

## SI7 Aharonov–Bohm interference in electron-doped odd- and even-denominator FQH states

Figs. S7a-e show AB pajamas at constant filling for $\nu = 1$ and the inner modes of $\nu = \frac{4}{3}, \frac{3}{2}, \frac{5}{3}, 2$. For all filling factors, clear AB-dominated interference patterns with negative slopes of the constant-phase lines are observed, consistent with expectations for electron doping. Figs. S7f-j present the corresponding 2D-FFTs of the data shown in Figs. S7a-e, respectively, plotted as a function of $\frac{\Phi_0}{\Delta B}$ and $\frac{1}{\Delta V_{\text{PG}}}$. For the integer filling factors $\nu = 1$ and 2, where the interfering charge is expected to be $e$, the interfering area $A$ is extracted from the 2D-FFTs shown in Figs. S7f and S7j. This provides upper and lower bounds for $A$ for intermediate filling factors between $1 < \nu < 2$. Averaging these bounds yields an interfering area of $A = 0.99 \pm 0.10$ $\mu\text{m}^2$.

To determine the magnetic field periodicity $\Delta B$, we proceed as in Sec. SI6. Figs. S7k-o show 1D-FFT amplitudes as a function of $\frac{\Phi_0}{\Delta B}$, taken at the peak value of $\Delta V_{\text{PG}}$. By fitting a Gaussian curve to the data, we extract the mean value $\frac{\Phi_0}{\Delta B_{\text{fit}}}$ and the variance $\sigma^2$, as shown in the lower-right sides of Figs. S7k-o.

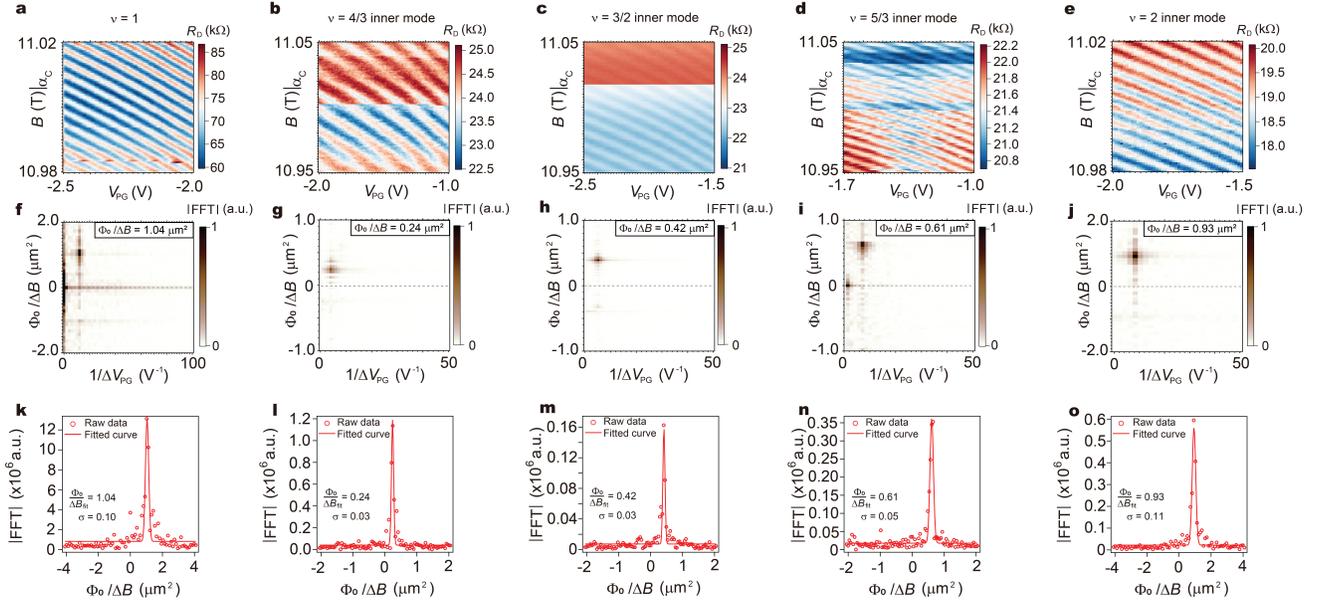

**Figure S7: Electron-doped Aharonov–Bohm interference.** (**a-e**) $R_D$ for $\nu = 1$, $\frac{4}{3}$ inner mode, $\frac{3}{2}$ inner mode, $\frac{5}{3}$ inner mode, and 2 inner mode, measured in $B|_{\alpha_c}$-$V_{\text{PG}}$ plane. (**f-j**) 2D-FFT as a function of $\frac{\Phi_0}{\Delta B}$ and $\frac{1}{\Delta V_{\text{PG}}}$, corresponding to Figs. S7a-e, respectively. (**k-o**) 1D cuts of the 2D-FFT as a function of $\frac{\Phi_0}{\Delta B}$, taken at the peak value of $\Delta V_{\text{PG}}$, obtained from Figs. S7f-j, respectively. Lower-left sides: Gaussian fitting parameters.

## SI8 Temperature dependence on resistance oscillations for $\nu = -\frac{1}{2}$

Here, we show $\Delta R_\mathrm{D}$ for $\nu = -\frac{1}{2}$, measured in the $B|_{\alpha_c}$-$V_\mathrm{PG}$ plane at temperatures of 10 mK, 50 mK, 80 mK, and 100 mK. As expected, the visibility decreases with increasing temperature due to the dephasing of the interfering quasiparticle. The magnetic-field periodicity, shown in the insets of Figs. S8a-d, does not change over the entire temperature range.

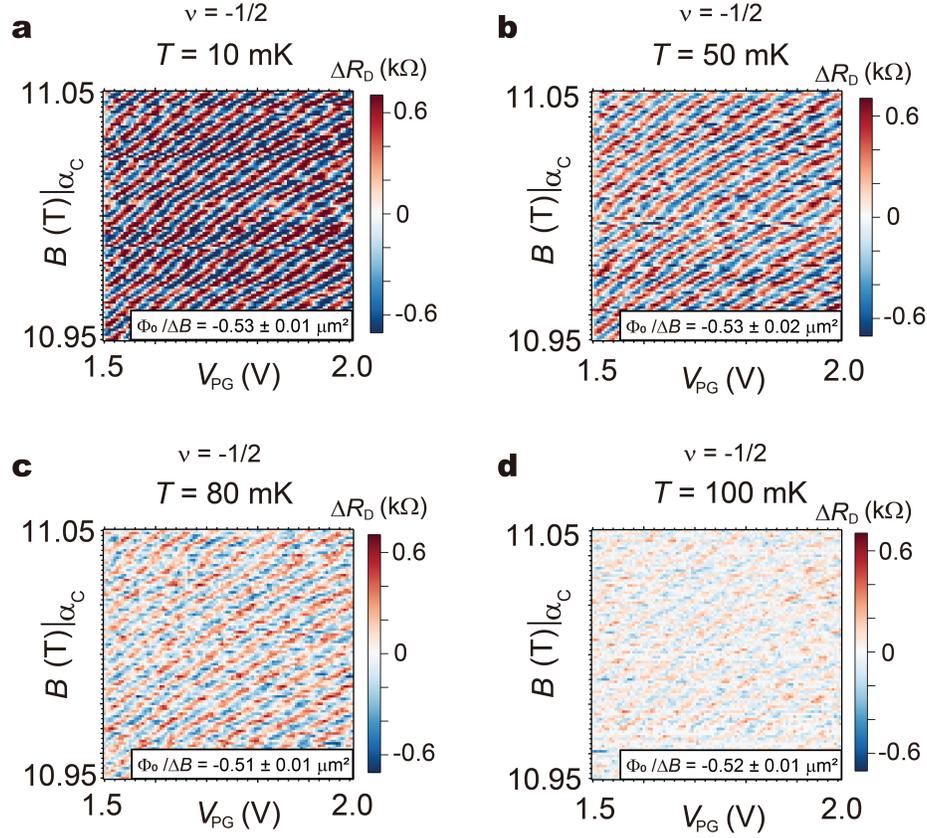

**Figure S8: Temperature dependent Aharonov-Bohm interference for $\nu = -\frac{1}{2}$.** (**a-d**) $\Delta R_\mathrm{D}$ at $\nu = -\frac{1}{2}$ displayed in $B|_{\alpha_c}$-$V_\mathrm{PG}$ plane for different temperatures, (a) $T$ = 10 mK, (b) $T$ = 50mK, (c) $T$ = 80 mK, and (d) $T$ = 100 mK. $\alpha_c$ = -50 T/V is obtained using $C_\mathrm{RG}$ calculated in Sec. SI2

12

## SI9 Temperature dependence on resistance oscillations for $\nu = \frac{3}{2}$

As in the previous section, we present $\Delta R_D$ measurements for $\nu = \frac{3}{2}$, focusing on the interfering fractional inner edge mode attemperatures of 10 mK, 20 mK, 30 mK, and 50 mK. For these measurements, an $\alpha_c$ value of 15.3 T/V is chosen, resulting in $\Delta \Phi = (2.01 \pm 0.35)\Phi_0 \approx 2\Phi_0$ at 10 mK. As shown in Figs. S9a-d, the magnetic-field periodicity remains constant over this temperature range. However, in this case, visibility is lost above the lower temperature of 50 mK.

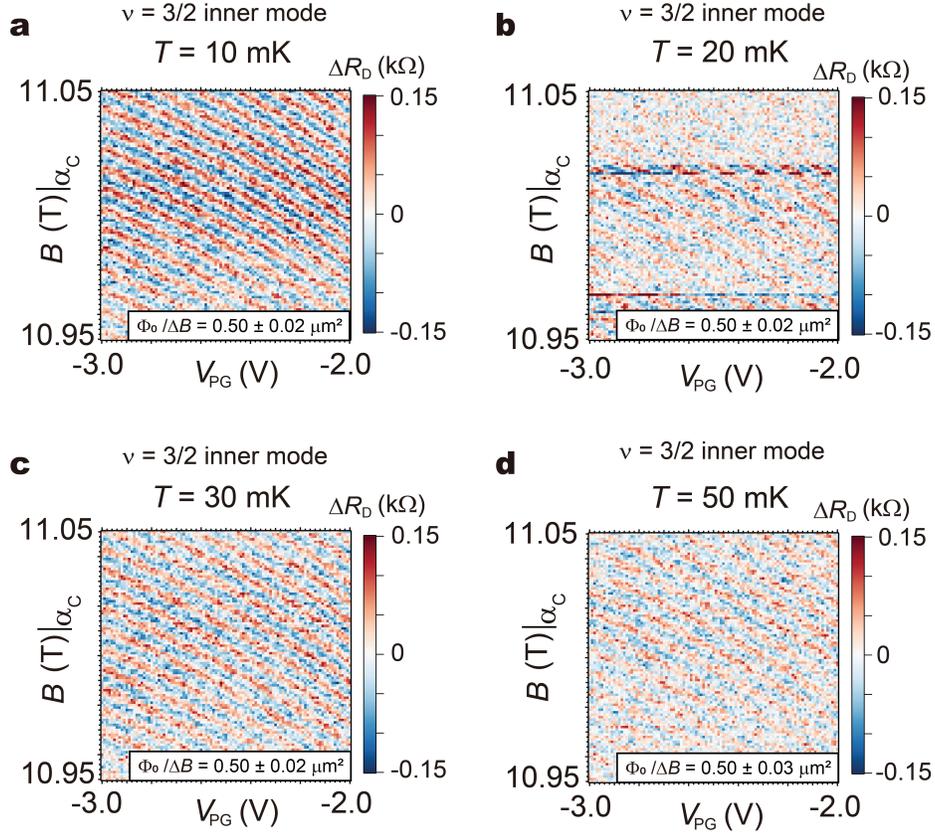

**Figure S9: Temperature dependent Aharonov-Bohm interference for $\nu = \frac{3}{2}$ inner mode.** (a-d) $\Delta R_D$ at $\nu = \frac{3}{2}$ inner mode displayed in $B|_{\alpha_c}$-$V_{PG}$ plane for different temperatures, (a) $T$ = 10 mK, (b) $T$ = 20mK, (c) $T$ = 30 mK, and (d) $T$ = 50 mK. $\alpha_c$ = 15.3 T/V is obtained using $C_{CG}$ calculated in Sec. SI12.

## SI10 Aharonov-Bohm interference for $\nu = \frac{3}{2}$ outer mode

At the bulk filling $\nu = \frac{3}{2}$, we also measured interference of the integer outer mode. The AB pajama in Fig. S10a and its 2D-FFT in Fig. S10b yield $\frac{\Phi_0}{\Delta B} = 1.11$ $\mu m^2$, consistent with an interfering charge $e$. It also indicates that the interference area is slightly larger than for the inner-mode interference, as expected.

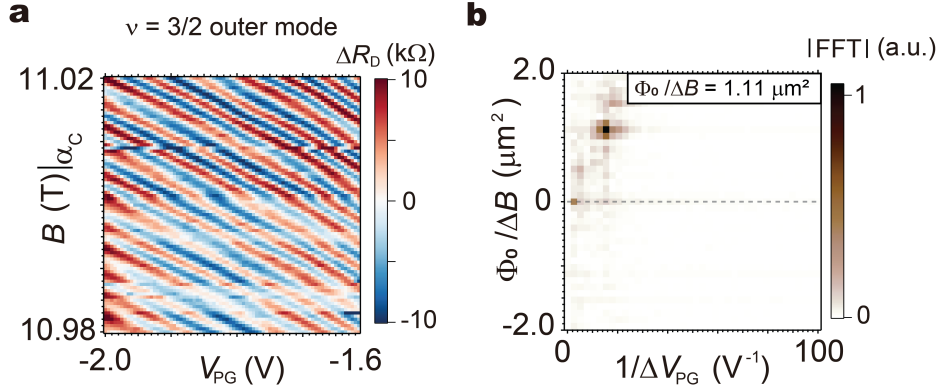

**Figure S10: Aharonov-Bohm interference for the outer mode of $\nu = \frac{3}{2}$.** (a) $\Delta R_D$ at $\nu = \frac{3}{2}$ displayed in $B|_{\alpha_c}$-$V_{PG}$ plane, with partitioning of the integer outer mode. (b) Corresponding 2D-FFT analysis used to extract the magnetic field periodicity $\frac{\Phi_0}{\Delta B}$.

14

## SI11 Derivation of the statistical contribution to the magnetic-field periodicity

As discussed in the main text, the diagonal resistance $R_D = \frac{V_D^+ - V_D^-}{I_{SD}}$ includes an oscillatory contribution $\Delta R_D \propto \cos\theta$, with the interference phase $\theta$ as given in Eq. (1) in the main text. In particular, this phase depends on the number $N_2$ of localized quasiparticles inside the interference loop. It is plausible that, in the course of the measurement, this number fluctuates. In this section, we compute the average of the oscillatory term $\langle \Delta R_D \rangle$ over this number. The phase $\theta$ is given by

$$\theta = 2\pi \left( \frac{e_1^*}{e} \frac{AB}{\Phi_0} + \frac{\theta_{12}}{2\pi} N_2 \right) \tag{S4}$$

with $e_1^*$ the interfering quasiparticle charge, $e_2^*$ the charge of localized quasiparticles, and $\theta_{12}$ is the statistical phase of $e_1^*$ encircling $e_2^*$. The interferometer area $A$ is assumed constant throughout this analysis.

The average oscillatory term is given by

$$\langle \Delta R_D \rangle = \sum_{N_2} P(N_2) \cos\left[ 2\pi \left( \frac{e_1^*}{e} \frac{AB}{\Phi_0} + \frac{\theta_{12}}{2\pi} N_2 \right) \right], \tag{S5}$$

where $P(N_2)$ is the probability distribution of $N_2$. We will treat $N_2$ as a discrete random variable with a Gaussian distribution with mean $\langle N_2 \rangle$ and variance $\sigma$, i.e.,

$$P(N_2) = \mathcal{N} \exp\left[ -\frac{(N_2 - \langle N_2 \rangle)^2}{2\sigma^2} \right], \tag{S6}$$

with $\mathcal{N}$ a normalization constant. The average number of quasiparticles, $\langle N_2 \rangle$, depends continuously on the electron density and magnetic field and is not required to be an integer. To evaluate the average, we employ the Poisson summation formula to write

$$\langle \Delta R_D \rangle = \mathcal{N} \sum_s \int_{-\infty}^{\infty} dN_2 \exp\left[ -i2\pi s N_2 - \frac{(N_2 - \langle N_2 \rangle)^2}{2\sigma^2} \right] \cos\left[ 2\pi i \left( \frac{e_1^*}{e} \frac{AB}{\Phi_0} + \frac{\theta_{12}}{2\pi} N_2 \right) \right]. \tag{S7}$$

Performing the Gaussian integral over $N_2$, we obtain

$$\langle \Delta R_D \rangle = \sqrt{2\pi}\sigma \mathcal{N} \sum_s e^{-\frac{\sigma^2}{2}(\theta_{12} - 2\pi s)^2} \cos\left[ 2\pi \frac{e_1^*}{e} \frac{AB}{\Phi_0} + (\theta_{12} - 2\pi s)\langle N_2 \rangle \right]. \tag{S8}$$

For $\sigma^2 \gtrsim 1$, the sum is dominated by the integers that minimize $(\theta_{12} - 2\pi s)^2$. We begin with $\theta_{12} \neq (2n+1)\pi$, for which there is a single dominant integer $s_0$. Keeping only the corresponding term we obtain

$$\langle \Delta R_D \rangle \approx e^{-\frac{\sigma^2}{2}(\theta_{12} - 2\pi s)^2} \cos\left[ 2\pi \frac{e_1^*}{e} \frac{AB}{\Phi_0} + (\theta_{12} - 2\pi s_0)\langle N_2 \rangle \right]. \tag{S9}$$

Next, we compute the average $\langle N_2 \rangle$. We define $\nu$ as the 'pristine' filling factor, i.e., the rational number defining the FQH plateau. When the magnetic field $B$ and electron number $N$ satisfy $N = \nu \frac{AB}{\Phi_0}$,

15

there are no quasiparticles in the ground state. Deviations from the relation via $B \to B + \Delta B$ or $N \to N + \Delta N$ introduce a number of charge $e_2^*$ quasiparticles given by

$$e_2^* \langle N_2 \rangle = e \Delta N + \frac{e \nu A}{\Phi_0} \Delta B . \tag{S10}$$

The first term on the right-hand side corresponds to the charge entering the interfering loop due to the change in $V_{CG}$, i.e., $e \Delta N = C A \Delta V_{CG}$ with $C$ the CG capacitance per unit area. The second term encodes excitations that arise as the change in the magnetic field modifies the Landau-level degeneracy.

In the experiment, the interference pattern was recorded along specific lines in the $B$–$V_{CG}$ plane parameterized by $\alpha = \frac{\partial B}{\partial V_{CG}}$. Along such trajectories, $\Delta N$ and $\Delta B$ are related via $\Delta N = \frac{CA}{e\alpha} \Delta B$. and the average number of quasiparticles is thus given by

$$\langle N_2 \rangle = \frac{e}{e_2^*} \left( \frac{CA}{e} \frac{1}{\alpha} - \frac{\nu A}{\Phi_0} \right) \Delta B . \tag{S11}$$

Inserting this average into Eq. (S9) we finally obtain, up to a constant phase inside the cosine,

$$\langle \Delta R_D \rangle \approx e^{-\frac{\sigma^2}{2}(\theta_{12} - 2\pi s)^2} \cos\left[ \left( 2\pi \frac{e_1^*}{e} + (\theta_{12} - 2\pi s_0) \frac{\nu e}{e_2^*} \frac{\alpha_c - \alpha}{\alpha} \right) \frac{A \Delta B}{\Phi_0} \right], \tag{S12}$$

where we used $\alpha_c = \frac{C \Phi_0}{\nu e}$. For $\alpha = \alpha_c$ we recover the pristine AB interference without quasiparticle contributions.

For the filling factors $\nu = \frac{4}{3}$ and $\frac{5}{3}$ with an interfering quasiparticle charge of $e_1^* = \nu_{LL} e$, we find a different $\Delta B$ dependence for bulk quasiparticles $e_2^* = \frac{1}{3} e$ or $e_2^* = \frac{2}{3} e$. The statistical phases for the two types of bulk quasiparticles are $\theta_{12} = \frac{2\pi}{3}$ (such that $s_0 = 0$) and $\frac{4\pi}{3}$ (such that $s_0 = -1$), respectively, i.e.,

$$e_2^* = \frac{1}{3} : \langle \Delta R_D \rangle_{\nu = \frac{4}{3}, \frac{5}{3}} \approx e^{-\frac{2\sigma^2 \pi^2}{9}} \cos\left[ 2\pi \left( \nu_{LL} - \nu + \nu \frac{\alpha_c}{\alpha} \right) \frac{A \Delta B}{\Phi_0} \right] \tag{S13}$$

$$e_2^* = \frac{2}{3} : \langle \Delta R_D \rangle_{\nu = \frac{4}{3}, \frac{5}{3}} \approx e^{-\frac{2\sigma^2 \pi^2}{9}} \cos\left[ 2\pi \left( \nu_{LL} - \nu - \nu \frac{\alpha_c}{\alpha} \right) \frac{A \Delta B}{\Phi_0} \right] . \tag{S14}$$

At $\nu = \frac{3}{2}$ we again take $e_1^* = \nu_{LL} e$. For the bulk quasiparticles, we consider $e_2^* = \frac{1}{2}$ in which case $\theta_{12} = \frac{\pi}{2}$ or $e_2^* = \frac{1}{4}$, for which $\theta_{12} = \pi$. In the first case, the sum in Eq. (S8) is dominated by $s_0 = 0$, but in the second case, $s_0 = 0$ and $s_0 = 1$ contribute equally. Proceeding as before, we find

$$e_2^* = \frac{1}{2} : \langle \Delta R_D \rangle_{\nu = \frac{3}{2}} \approx 2 e^{-\frac{\sigma^2 \pi^2}{2}} \cos\left[ 2\pi \nu_{LL} \frac{A \Delta B}{\Phi_0} \right] \cos\left[ 2\pi \left( -\nu + \nu \frac{\alpha_c}{\alpha} \right) \frac{A \Delta B}{\Phi_0} \right] , \tag{S15}$$

$$e_2^* = \frac{1}{4} : \langle \Delta R_D \rangle_{\nu = \frac{3}{2}} \approx e^{-\frac{\sigma^2 \pi^2}{8}} \cos\left[ 2\pi \left( \nu_{LL} - \nu + \nu \frac{\alpha_c}{\alpha} \right) \frac{A \Delta B}{\Phi_0} \right] . \tag{S16}$$

16

We note that double windings of the fundamental charge $e_1^* = \frac{1}{3}e$ or $\frac{1}{4}e$ would yield the same phase evolution as a single winding of twice that charge, and this analysis cannot distinguish between these two scenarios. In contrast, the charge of the localized quasiparticles, $e_2^*$, can be inferred by examining the slope of $\frac{\Phi_0}{\Delta B}$ as a function of $\frac{1}{\alpha}$. The data for $\nu = \frac{3}{2}, \frac{4}{3}, \frac{5}{3}$ (main text), and for $\nu = -\frac{1}{2}, -\frac{2}{3}$ (Fig. S11) follow the slope $\nu \alpha A$. In particular, at both half-filled states, there is a single periodicity, which is inconsistent with Eq. (S15). We conclude that in all the measured states, deviations of the filling factor introduce bulk quasiparticles with the fundamental charge.

### SI12 Determination of the capacitance of the center gate from $\alpha$-dependent AB interference

Fig. S11 summarizes the $\alpha$-dependence of $\frac{\Phi_0}{\Delta B}$ for all fractional fillings in the current study as well as $\nu = \frac{1}{3}$ from our earlier experiment.[7] These results provide information on the CG capacitance. The constant-phase lines of the cosine in Eq. (S12) can be expressed as

$$\frac{\Phi_0}{\Delta B} = A\left(\nu_{LL} - \frac{e}{e_2^*}\frac{\tilde{\theta}_{12}}{2\pi}\nu\right) + A\left(\frac{e}{e_2^*}\frac{\tilde{\theta}_{12}}{2\pi}\frac{C_{CG}\Phi_0}{e}\right)\frac{1}{\alpha}, \tag{S17}$$

where we used $e_1^* = \nu_{LL} e$, $\tilde{\theta}_{12} = \theta_{12} - s_0$, and kept the dependence on the capacitance explicit. In particular, we use the CG capacitance $C_{CG}$, which can be different from the RG capacitance $C_{RG}$, denoted by $C$ in the main text. The first term ('bias term') on the right-hand side of Eq. (S17) is independent of the capacitance. When $e_2^*$ is the fundamental quasiparticle charge and bulk-edge couplings are absent, the factor $\frac{e}{e^*}\frac{\tilde{\theta}_{12}}{2\pi} = 1$. The intercept of $\frac{\Phi_0}{\Delta B}$ with the $\frac{1}{\alpha} = 0$ line in Fig. S11 thus directly quantifies how much $\tilde{\theta}_{12}$ deviates from the ideal case. The second term ('slope term') on the right-hand side of Eq. (S17) is proportional to $C_{CG}$. Fitting the measured data shown in Fig. S11 to a line thus determines both $\tilde{\theta}_{12}$ and $C_{CG}$

Table S1 summarizes the results of fitting the data for each filling factor. The significantly smaller value of $C_{CG}$ for $\nu = \frac{1}{3}$ from the previous study can be explained by the different thicknesses of the top hBN, which was 48 nm for the previous study and 29 nm for the current study.

In Sec. SI2, we calculated $C_{RG}$ = 0.959 mF/m$^2$ from the Landau fan diagram, which deviates from the $C_{CG}$ values in Table S1 by 10% at $\nu = \frac{4}{3}$ and less than 7% at the other fillings. The actual difference between $C_{RG}$ and $C_{CG}$ is thus about half the value of 15% quoted in the main text, which did not take into account deviations $\tilde{\theta}_{12}$ from the theoretically expected value.

17

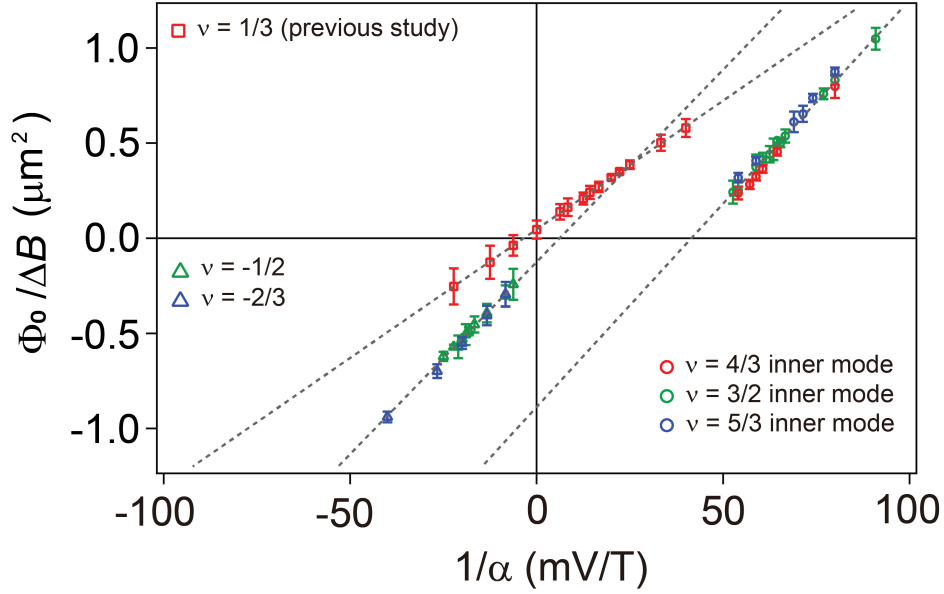

**Figure S11: The statistical contribution to the interference of fractional quasiparticles.** The magnetic field periodicity $\frac{\Phi_0}{\Delta B}$ obtained from 2D-FFTs is plotted as a function of $\frac{1}{\alpha}$ for the various fractional modes in the current study and the previous study.[7]

**Table S1: Determination of the braiding phase $\tilde{\theta}_{12}$ and the capacitance of the center gate $C_{\text{CG}}$.**

| partitioned edge | $A$ (μm²) | bias term (μm²) | slope term (Tμm²/mV) | $\frac{e}{e^*_z} - \frac{\tilde{\theta}_{12}}{2\pi}$ | $C_{\text{CG}}$ (mF/m²) |
|---|---|---|---|---|---|
| $\nu$ = 4/3 inner mode | 0.99 ± 0.10 (see SI7) | -0.967 ± 0.027 | 22.04 ± 0.43 | 0.98 ± 0.02 | 0.876 ± 0.017 |
| $\nu$ = 3/2 inner mode | | -0.887 ± 0.019 | 21.38 ± 0.28 | 0.93 ± 0.01 | 0.898 ± 0.011 |
| $\nu$ = 5/3 inner mode | | -0.843 ± 0.052 | 21.26 ± 0.77 | 0.91 ± 0.05 | 0.912 ± 0.033 |
| $\nu$ = -1/2 | 1.02 ± 0.06 (see SI6) | -0.124 ± 0.009 | 20.21 ± 0.54 | 0.75 ± 0.01 | 1.013 ± 0.027 |
| $\nu$ = -2/3 | | -0.136 ± 0.014 | 20.40 ± 0.60 | 0.80 ± 0.01 | 0.967 ± 0.028 |
| $\nu$ = 1/3 (previous study) | 1.00 ± 0.12 | 0.046 ± 0.001 | 13.52 ± 0.07 | 0.86 ± 0.01 | 0.607 ± 0.003 |

**References for Supplementary Material**

1. Venugopal, A. *et al.* Effective mobility of single-layer graphene transistors as a function of channel dimensions. *Journal of Applied Physics* **109** (2011).

2. Li, J. *et al.* Even-denominator fractional quantum Hall states in bilayer graphene. *Science* **358**, 648–652 (2017).

3. Kumar, R. *et al.* Quarter-and half-filled quantum Hall states and their competing interactions in bilayer graphene. *arXiv preprint arXiv:2405.19405* (2024).

4. Nakamura, J. *et al.* Aharonov-Bohm interference of fractional quantum Hall edge modes. *Nature Physics* **15**, 563–569 (2019).

5. Nakamura, J., Liang, S., Gardner, G. C. & Manfra, M. J. Impact of bulk-edge coupling on observation of anyonic braiding statistics in quantum Hall interferometers. *Nature Communications* **13**, 344 (2022).

6. Nakamura, J., Liang, S., Gardner, G. C. & Manfra, M. J. Direct observation of anyonic braiding statistics. *Nature Physics* **16**, 931–936 (2020).

7. Kim, J. *et al.* Aharonov–Bohm interference and statistical phase-jump evolution in fractional quantum Hall states in bilayer graphene. *Nature Nanotechnology* 1–8 (2024).



\end{document}